\documentclass[journal]{IEEEtran}
\IEEEoverridecommandlockouts
\usepackage{cite}
\usepackage{amsmath,amssymb,amsfonts}
\usepackage{algorithmic}
\usepackage{graphicx}
\usepackage{textcomp}
\usepackage{xcolor}
\usepackage{pgfplots}
\usepackage{tikz}
\usepackage{siunitx}
\usepackage{graphicx}
\usepackage{threeparttable}
\usepackage{tabulary}
\usepackage{booktabs}
\usepackage{footnote}
\usepackage{multirow}
\usepackage{csvsimple}
\usepackage{svg}

\usepackage{xargs}                      
\usepackage[colorinlistoftodos,prependcaption,textsize=tiny]{todonotes}
\newcommandx{\unsure}[2][1=]{\todo[linecolor=red,backgroundcolor=red!25,bordercolor=red,#1]{#2}}
\newcommandx{\change}[2][1=]{\todo[linecolor=blue,backgroundcolor=blue!25,bordercolor=blue,#1]{#2}}
\newcommandx{\info}[2][1=]{\todo[linecolor=OliveGreen,backgroundcolor=OliveGreen!25,bordercolor=OliveGreen,#1]{#2}}
\newcommandx{\improvement}[2][1=]{\todo[linecolor=Plum,backgroundcolor=Plum!25,bordercolor=Plum,#1]{#2}}
\newcommandx{\thiswillnotshow}[2][1=]{\todo[disable,#1]{#2}}

\def\BibTeX{{\rm B\kern-.05em{\sc i\kern-.025em b}\kern-.08em
    T\kern-.1667em\lower.7ex\hbox{Ehttps://v2.overleaf.com/project/5b154d124aedb43cc5e200cc}\kern-.125emX}}

\renewcommand{\baselinestretch}{0.98}

\newcommand\copyrighttext{%
  \footnotesize This work has been submitted to the IEEE for possible publication. Copyright may be transferred without notice, after which this version may no longer be accessible.}
\newcommand\copyrightnotice{%
\begin{tikzpicture}[remember picture,overlay]
\node[anchor=south,yshift=10pt] at (current page.south) {\fbox{\parbox{\dimexpr\textwidth-\fboxsep-\fboxrule\relax}{\copyrighttext}}};
\end{tikzpicture}%
}

\begin{document}

\title{Always-On 674uW @ 4GOP/s Error Resilient Binary Neural Networks with Aggressive SRAM Voltage Scaling on a 22nm IoT End-Node}

\author{Alfio Di Mauro, Francesco Conti, Pasquale Davide Schiavone, Davide Rossi, Luca Benini
\thanks{
A. Di Mauro, F. Conti, P. D. Schiavone and L. Benini are with the Integrated Systems Laboratory, D-ITET, ETH Z\"urich, 8092 Z\"urich, Switzerland.
F. Conti, D. Rossi and L. Benini are also with the Energy-Efficient Embedded Systems Laboratory, DEI, University of Bologna, 40126 Bologna, Italy.
\protect\\}
}
\maketitle
\copyrightnotice

\begin{abstract}
Binary Neural Networks (BNNs) have been shown to be robust to random bit-level noise, making aggressive voltage scaling attractive as a power-saving technique for both logic and SRAMs.
In this work, we introduce the first fully programmable IoT end-node system-on-chip (SoC) capable of executing software-defined, hardware-accelerated BNNs at ultra-low voltage.
Our SoC exploits a hybrid memory scheme where error-vulnerable SRAMs are complemented by reliable standard-cell memories to safely store critical data under aggressive voltage scaling.
On a prototype in 22nm FDX technology, we demonstrate that both the logic and SRAM voltage can be dropped to 0.5V without any accuracy penalty on a BNN trained for the CIFAR-10 dataset, improving energy efficiency by 2.2X w.r.t. nominal conditions. Furthermore, we show that the supply voltage can be dropped to 0.42V (50\% of nominal) while keeping more than 99\% of the nominal accuracy (with a bit error rate $\sim$1/1000). In this operating point, \textcolor{black}{our prototype performs 4Gop/s (15.4 Inference/s on the CIFAR-10 dataset)} by computing up to \textcolor{black}{13} binary ops per pJ, \textcolor{black}{achieving 22.8 Inference/s/mW} while keeping within a peak power envelope of 674uW -- low enough to enable always-on operation in ultra-low power smart cameras, long-lifetime environmental sensors, and insect-sized pico-drones.
\end{abstract}


\section{Introduction}
\label{sec:intro}

\IEEEPARstart{T}{he} latest advances in the Internet-of-Things (IoT) are changing the nature of edge-computing devices. End-nodes have to support, in-place, an increasing range of functionality, for example, video and audio sensory data processing, and complex systems control strategies.
These new capabilities will enable applications such as an entirely new class of biomedical
devices~\cite{GreenspanGuestEditorialDeep2016}, autonomous insect-sized drones~\cite{Palossi64mWDNNbasedVisual2019}, cheap smart sensors~\cite{ManicIntelligentBuildingsFuture2016} to continuously check the stability of bridges, tunnels, and other buildings. Machine learning algorithms and specifically deep neural networks (DNNs) have shown outstanding performance in performing these tasks.
However, while DNNs fits well the performance and power budgets of embedded GPUs and FPGAs, deploying such \textcolor{black}{compute-intensive} algorithms on battery-powered IoT end-node platforms, characterized by heavily constrained power budgets (typically \SI{100}{\micro\watt} to \SI{100}{\milli\watt}), still constitutes a huge challenge, as they are expected to achieve lifetimes in the orders of months, years or even decades. As such, recent research efforts from both industry and academia have focused on enabling the deployment of deep inference on devices operating in the \SIrange{10}{100}{\milli\watt} power range~\cite{UNPU2019,BankmanAlwaysOn8mJ86,BiswasConvRAMEnergyEfficientSRAM,BahouXNORBIN95TOp2018,FlamandGAP8RISCVSoC2018,XuScalingedgeinference2018,SharifyLoomExploitingWeight2018,ChenEyerissEnergyEfficientReconfigurable2017}.

The most common approach to reducing average power consumption, widely used in commercial microcontrollers and IoT end-nodes, is duty cycling (a.k.a. sleep-walking). According to this paradigm, the system stays in deep-sleep mode for most of the time, featuring a power consumption in the range of \SI{100}{\nano\watt} to \SI{10}{\micro\watt}, and wakes up to perform the acquisition and classification task (e.g. with a CNNs), only when it wakes up, either by an externally triggered event or by an internal timer. However, in the first case, this approach requires trigger fine-tuning to reduce the number of false-positive activations, which could be fairly difficult to achieve in a real scenario, and which would feature poor generalization capabilities in other contexts. Alternatively, if a time-based trigger is employed, therefore exploiting a simple triggering mechanism, the computing cost of an accurate CNNs is so high that the active energy becomes rapidly dominant even at low duty cycling rates. As a result, the latter approach is inefficient whenever fast reaction time is required at the sensor edge.

A common technique to reduce the active power of deeply embedded computing platforms is near-threshold computing \cite{Dreslinski}. Scaling voltage together with frequency allows improving significantly the energy efficiency of computations, by exploiting the quadratic dependency of dynamic power with supply voltage.
However, aggressive voltage scaling has a significant impact on the operating frequency of the logic, and on the reliability of the memory elements of the system, especially those based on SRAMs. While the frequency degradation at low voltage can be recovered by exploiting powerful and efficient hardware accelerators, the SRAM reliability issue remains an unsolved problem. Therefore, in most cases 6T-SRAMs have to be replaced by more resilient, custom solutions such as SRAMs composed of 8T or 10T bitcells supported by reading and writing assist circuits~\cite{Verma65nm8TSubVt2007,verma2008sram8t}. Among the approaches adopted to improve the resiliency of memory elements at low voltage, usage of standard cell memories is particularly convenient since they are built on top of standard library cells such as flip-flop or latches, much more resilient than SRAMs when operating close to the threshold voltage of transistors \cite{teman_controlled_2015,alioto-latchmem}. This comes with a significant cost in terms of area. On the other hand, a relatively large on-chip memory is necessary to enable complex algorithms based on DNNs~\cite{SzeEfficientProcessingDeep2017,XuScalingedgeinference2018}.  

\textcolor{black}{In the last years, BNNs  ~\cite{RastegariXNORNetImageNetClassification2016a,HubaraQuantizedNeuralNetworks2016,hubara2016}} became popular in the embedded computing domain for having achieved remarkable accuracy on many complex classification tasks, narrowing the gap that separates them from the state-of-the-art fixed point or floating point CNNs.
Compared to fixed or floating point CNN implementation, which relies on convolutions, BNNs are characterized by a very lightweight hardware implementation of the data path. Binary convolution can be implemented with simple logic elements such as XNOR gates, requiring a very limited amount of area-hungry adders for partial sum accumulation. Moreover, BNNs also feature lower memory footprints compared to CNNs, reducing the amount of energy consumed on the memory side for weights and intermediate results storage.

Those features make BNNs a good candidate for all those scenarios where power consumption is a major concern, but at the same time, a high responsiveness to the sensor stimuli needs to be ensured (e.g. pico-sized autonomous navigation robots or surveillance nodes). The low power envelope that characterizes BNNs allows to use them as data pre-filtering algorithms, specifically, to extract semantically meaningful information in an always-on operating mode~\cite{Yangalwayson}. In this context, BNNs can be used as a first inference stage of a staged inference pipeline, composed by low-power, less-accurate early inference stages, and computationally-powerful fixed or floating point CNN implementations as latter stages~\cite{Verhelst-cascaded}. As the memory footprint of BNNs is significantly lower than CNNs, bigger topologies can be supported in the same power/performance budget, enhancing the generalization capability of the early stages, thereby lowering the false positive triggering occurrence. 
Additionally, the employment of BNNs as preliminary filtering stage does not prevent the adoption of conventional power-saving strategies like duty-cycling or sleep-walking at run time, whenever the application latency requirements allow it.

One of the advantages of DNNs, as well as BNNs, is the high robustness to noise \cite{Yang2018}.  The high resiliency of BNNs to random errors is given by the fact that as opposed to traditional neural networks, where activations and weights are represented by integer numbers, no bit in their activations and weights is inherently more significant than any other. As a consequence, no bit is more vulnerable than any other: information processing is spread equally among all bits, and only a very high error rate can bring a dramatic loss in quality-of-results.
The BNNs noise robustness is a very powerful feature since it enables very aggressive power reduction techniques to be applied also on the memories.

In this work, we advance the state-of-the-art with regards to ultra-low power deep inference with BNNs with three key contributions:

i) We propose a strategy to execute noisy BNNs on microcontrollers. To the best of our knowledge, in this work, we propose the first complete and fully programmable end node SoC architecture and BNN inference data and code allocation strategy enabling the execution of hardware accelerated BNNs at ultra-low voltage. \\
ii) We describe and demonstrate on silicon a hybrid memory architecture composed of big SRAMs for error-resilient data and smaller (Standard Cell Memories) SCMs to hold vulnerable data such as microcontroller instructions and stacks. In this work, we also provide a methodology to efficiently exploit such memory architecture. The hybrid memory architecture template presented in this work is easily applicable to other platforms. \\
iii) We present a self-test strategy for Bit Error Rate measurement performed on large SRAM. This approach allows characterizing SRAM memories at ultra-low voltages, thereby estimating the amount of noise injected in the BNN. \\
iv) We demonstrate the validity of this architectural concept on an advanced prototype manufactured in GlobalFoundries 22nm FDX technology, using the safe SCMs to hold a microcontroller program testing SRAM bit error rates with millions of random reads/writes, operating down to 420 mV (50\% of the nominal supply voltage) for both logic and memories.


Finally, we show that using the embedded hardware accelerator for BNNs, our prototype can be operated at \SI{18}{\mega\hertz} down-scaling voltage to \SI{420}{\milli\volt} for both logic and memories. In this operating point, the prototype achieves up to 99\% of the nominal accuracy on a BNN trained for the CIFAR-10 dataset, while operating with an energy efficiency of 170fJ/op and within a power envelope of \SI{674}{\micro\watt} -- enabling embedding of advanced BNN-based cognitive functionality in ultra-low power "TinyML" devices such as biomedical sensors, long-lifetime environmental sensors, and insect-sized pico-UAVs.

The rest of the paper is organized as follows: Section~\ref{sec:related} discusses other works in the state-of-the-art related to this proposed work. Section~\ref{sec:archi} introduces the proposed SoC architecture. Section~\ref{sec:resilience_analysis} discusses the simulations we performed to evaluate the resilience of BNNs against SRAM errors. Section~\ref{sec:results} details the experimental methodology used to evaluate the SoC and the results of the evaluation in terms of Bit Error Rate (BER), power and energy efficiency. Section~\ref{sec:conclusion} draws conclusions.

\section{Related Work}
\label{sec:related}

\textcolor{black}{Recently, there has been a strong push towards the deployment of sophisticated artificial intelligence (in particular DNNs) on tiny end-node architectures dedicated to the extreme edge of the IoT -- fostering a fast-growing \textit{TinyML} research community~\cite{TinyML}, which has explored the field from two converging directions.
On the one hand, in the direction of shrinking DNN topology \cite{Book-tinyML}, reducing the amount~\cite{pruning} and numerical precision of network parameters~\cite{pact}, moving from floating point down to highly quantized numerical representations, e.g. 8 or 4 bits, and ultimately to BNNs~\cite{RastegariXNORNetImageNetClassification2016a}.
On the other hand, edge computing platforms are supporting this trend by becoming more and more specialized to efficiently execute machine learning workloads~\cite{fcontiFulmine,SzeEfficientProcessingDeep2017}.}
In this section, we focus on the latter research direction and describe research works related to the SoC proposed in this paper in a top-down fashion. We start from software-programmable architectures targeting the end-nodes of the IoT, go through specialized heterogeneous and error-resilient hardware architectures, and end with dedicated architectures for CNN inference exploiting extreme quantization and error resiliency. 

\subsection{IoT End-Node Architectures}
\label{sec:rel_iot}

A fundamental element of all IoT end-node architectures is software programmability, typically based on tiny microcontrollers with ARM Cortex-M class processors. Significant commercial examples of such micro-systems have been proposed by all major embedded systems vendors such as TI~\cite{_texas-1}, STMicroelectronics~\cite{_stmicroelectronics-1}, NXP~\cite{_nxp}, and Ambiq~\cite{_ambiq}. These systems feature aggressive sleep-walking capabilities thanks to sub-\SI{10}{\micro\watt} deep-sleep modes leading to an extremely small average power. On the other hand, current research in IoT end nodes is moving towards optimizing both active and sleep states exploiting near-threshold and sub-threshold operation. These techniques further improve the energy efficiency and reduce power consumption during computation~\cite{MyersISSCC2015} \cite{bol_sleepwalker_2013} \cite{AliotoApproxMem} \cite{RoyISQED2016} \cite{Independent-biasing-adimauro}.
Mr.~Wolf~\cite{PulliniMrWolfGFLOP2018} couples aggressive deep-sleep capabilities with an energy-proportional architecture, exceeding the computational capabilities of ULP microcontrollers by 2 orders of magnitude while offering a competitive energy efficiency also at low and sporadic workloads. This is achieved thanks to a heterogeneous parallel architecture composed of an always-on autonomous I/O subsystem, coupled with a parallel accelerator with 8 floating-point capable RISC-V cores.
To target specific computation domains such as CNNs, some commercial architectures leverage lightweight SW acceleration and optimized DSP libraries to improve performance. A well-known example is that of CMSIS developed by ARM, a set of libraries to optimize DSP applications on Cortex-M architectures, and CMSIS-NN~\cite{LaiCMSISNNEfficientNeural2018}, tuned to the deployment of embedded neural networks. An extension to these libraries has been proposed by Rusci~et~al.~\cite{RusciWorkinProgressQuantizedNNs2018} targeting highly quantized networks such as 4-bit, 2-bit and binarized networks \cite{garofalo2019pulpnn}.

However, due to their 32-bit nature, fully programmable solutions can only partially exploit the benefits of quantized NNs. While this approach significantly reduces the memory footprint of CNN, several additional operations are required to pack/unpack activation and weights to arithmetic formats suitable for software processing (e.g. 16-bit or 8-bit)~\cite{RusciWorkinProgressQuantizedNNs2018}, degrading performance and energy efficiency of inference. \textcolor{black}{Modern microcontrollers introduced dedicated ISA extensions to efficiently perform sub-word, sub-byte and SIMD operations \cite{RISCY,ARM-HELIUM}, and mitigate such performance degradation.}

To improve the overall efficiency of systems dedicated to NN acceleration, recent SoCs couple programmable processors with hardwired accelerators, in some cases exploiting low-precision functional units to exploit resiliency of CNNs to quantization.
Intel presented an IoT edge mote integrating an x86 processor accelerated by dedicated functional units for CNN cryptography workloads ~\cite{INTELISSCC2018}.
Conti~et~al. proposed Fulmine~\cite{fcontiFulmine}, a heterogeneous SoC coupling four general-purpose processors with a convolutional accelerator. While convolutional layers of CNNs run on the accelerator, other functions such as activations and pooling execute on the software processing cluster.
GAP-8~\cite{GAP8} includes a specialized accelerator for convolutional neural networks supporting 16-bit precision for activations and 16-bit, 8-bit and 4-bit precision for weights, achieving up to 600~GMAC/s/W within \SI{75}{\milli\watt} of power envelope.
\textcolor{black}{Another notable device is the low-power vision sensor node presented by Qualcomm Technologies \cite{QUALCOMM-Platform}, which performs end-to-end always-on visual detection tasks thanks to an ultra-low power QVGA CMOS sensor and a full digital processor subsystem integrated as a single device. Such architecture allows to perform video processing at the sensor edge in a \SI{1}{\milli \watt} power envelope, exploiting low-resolution sensing, data sparsity and event-driven computing, ultimately outputting only meta-information when meaningful events are detected.}

In this work we propose a near-threshold SoC joining the flexibility of a software programmable 32-bit RISC-V processor integrated into a state-of-the-art microcontroller featuring a rich set of peripherals, with the performance boost of a dedicated accelerator for BNN workloads, pushing quantization to the limit. On top of the flexibility and performance of this heterogeneous architecture, in this paper, we propose a heterogeneous memory architecture exploiting the error resiliency of BNN with respect to random errors in the memory system. To our best knowledge, the SoC described in this paper reports the lowest full system power for active operation and always-on BNN inference presented in industry or academia. 
\subsection{Heterogeneous and Error Resilient Memory Architectures}
\label{sec:rel_mem}

Optimizing the memory hierarchy is one of the main concerns in IoT end-nodes operating in near-threshold, since memory can be the dominant source of power consumption, potentially jeopardizing their energy efficiency ~\cite{bol_sleepwalker_2013,SchiavoneQuentinUltraLowPowerPULPissimo2018,ROSSI2016170}. While many approaches rely on the custom design of low-voltage memories ~\cite{calhoun2007sram8t,verma2008sram8t}, which come with the associated area and power overheads (e.g. 8T or 10T bitcells, read and write assist circuits)~\cite{Verma65nm8TSubVt2007}, an emerging trend relies on approximate SRAMs, often joined with precision/performance tunability or heterogeneous memory architectures. Frustaci et al. \cite{alioto-approx-SRAM} proposed an approximated SRAMs for error-tolerant applications, in which energy is saved at the cost of the occurrence of read/write errors by exploiting voltage scaling, selective error correction code (SECC), and selective write assist techniques (SNBB). Compared to the voltage scaling at iso-quality, the joint adoption of these techniques can provide more than $2\times $ energy reduction at a negligible area penalty. Other works propose the adoption of emerging technologies to realize approximate memory cells, such as RRAM \cite{Li-RRAM} and memristors \cite{Li-memristors}.

Although all the aforementioned approaches are effective, they all require the design of custom SRAM banks (either approximated or not), they feature deep circuit-level optimizations, that cannot be easily integrated into automatic memory generators. Other approaches exploit heterogeneous memory architectures mixing standard SRAMs and latch-based Standard Cell Memories (SCM). While SRAMs can not be considered reliable at relatively high voltages (e.g. 0.8V in the technology considered in this work), SCMs can operate in the same operating range of the rest of the logic, typically much wider \cite{teman_controlled_2015}. Tagliavini et al. \cite{TagliaviniSynergisticArchitectureProgramming2015}, proposed an HW/SW methodology to design energy-efficient ULP systems which combine the key ideas of a hybrid memory design where part of the memory system is approximated and part is precise, with an error-aware allocation strategy. Similarly to this work, our approach leverages standard 6T-SRAM cells that can be realized with memory generators provided by silicon vendors, and SCM that can be implemented with standard semi-custom design flows relying on industrially qualified standard-cells for implementation. On the other hand, our work exploits resiliency of binarized neural networks, where the position of the flip-bit error within the words is irrelevant to the quality of the final result, making them a much more suitable candidate for approximate computing.

\subsection{Dedicated Hardware Accelerators for DNNs and BNNs}
\label{sec:rel_hw}
Many dedicated hardware accelerators specifically designed to bring deep learning at an ultra-low power budget have been proposed.
Most designs employ fixed-point representation for weights and activations (e.g. Orlando~\cite{Desoli2017} achieving up to 2.9 Top/s/W).
Pruning and compression are popular techniques to further reduce the power budget~\cite{MoonsEnergyEfficientPrecisionScalableConvNet2017,Aimar2018,Han2016,Yuan2018,Jinshan2019}.

Binary Neural Networks~\cite{HubaraQuantizedNeuralNetworks2016,RastegariXNORNetImageNetClassification2016a} constitute a particularly interesting niche application due to their properties, as they can be trained to achieve similar results to full-precision counterparts~\cite{LinAccurateBinaryConvolutional2017} while keeping a smaller footprint, a more scalable structure and a higher resilience to errors, as further explored in Section~\ref{sec:resilience_analysis}.
\textit{FINN}~\cite{UmurogluFINNFrameworkFast2017},   was the first architecture capable of reaching more than 200 Gop/s/W on an FPGA.
Many of the most recent efforts towards the deployment of BNNs on silicon, such as \textit{BRein}~\cite{Ando2018}, \textit{XNOR-POP}~\cite{JiangXNORPOPprocessinginmemoryarchitecture2017a}, \textit{Conv-RAM}~\cite{BiswasConvRAMEnergyEfficientSRAM},
as well as the BTNN accelerator proposed by Yin~et~al.~\cite{Liu2019}, and the BNN accelerator presented by Wang~et~al.~\cite{Wang2018},
and Khwa~et~al.~\cite{Khwa65nm4KbAlgorithmDependent} have achieved an energy efficiency in the range of 10-50~TOP/s/W using in-memory computing.
Similar results have been claimed by more ``traditional'' ASICs such as \textit{UNPU}~\cite{UNPU2019} and \textit{XNORBIN}~\cite{Bahou2018}.
Mixed-signal ~\cite{BankmanAlwaysOn8mJ86}, \textcolor{black}{and in-memory mixed-signal approaches ~\cite{verma-In-Memory-Computing-CNN-Accelerator,verma-An-In-memory-Computing-DNN,Murmann-in-memory-mixed,Roy2019} are able to achieve up to 10-100$\times$ higher efficiency}, but paying a very significant cost in terms of design time, verification, and scalability to real systems.
Yang~et~al.~\cite{Yang2018} exploits one such system in their work, where similarly to what we propose SRAM is aggressively voltage-scaled to achieve a power benefit.

Our own system exploits a similar technique to Yang's, with the important distinction that their work is an extremely specialized ASIC, capable of executing a single BNN topology.
Rather, our design is a complete and fully programmable IoT end node on the line of those discussed in Section~\ref{sec:rel_iot}, augmented with a very small hardware accelerator~\cite{ContiXNORNeuralEngine2018}.

\section{Quentin SoC}
\label{sec:archi}

\begin{figure}[t!]
\centerline{\includegraphics[width=0.9\linewidth]{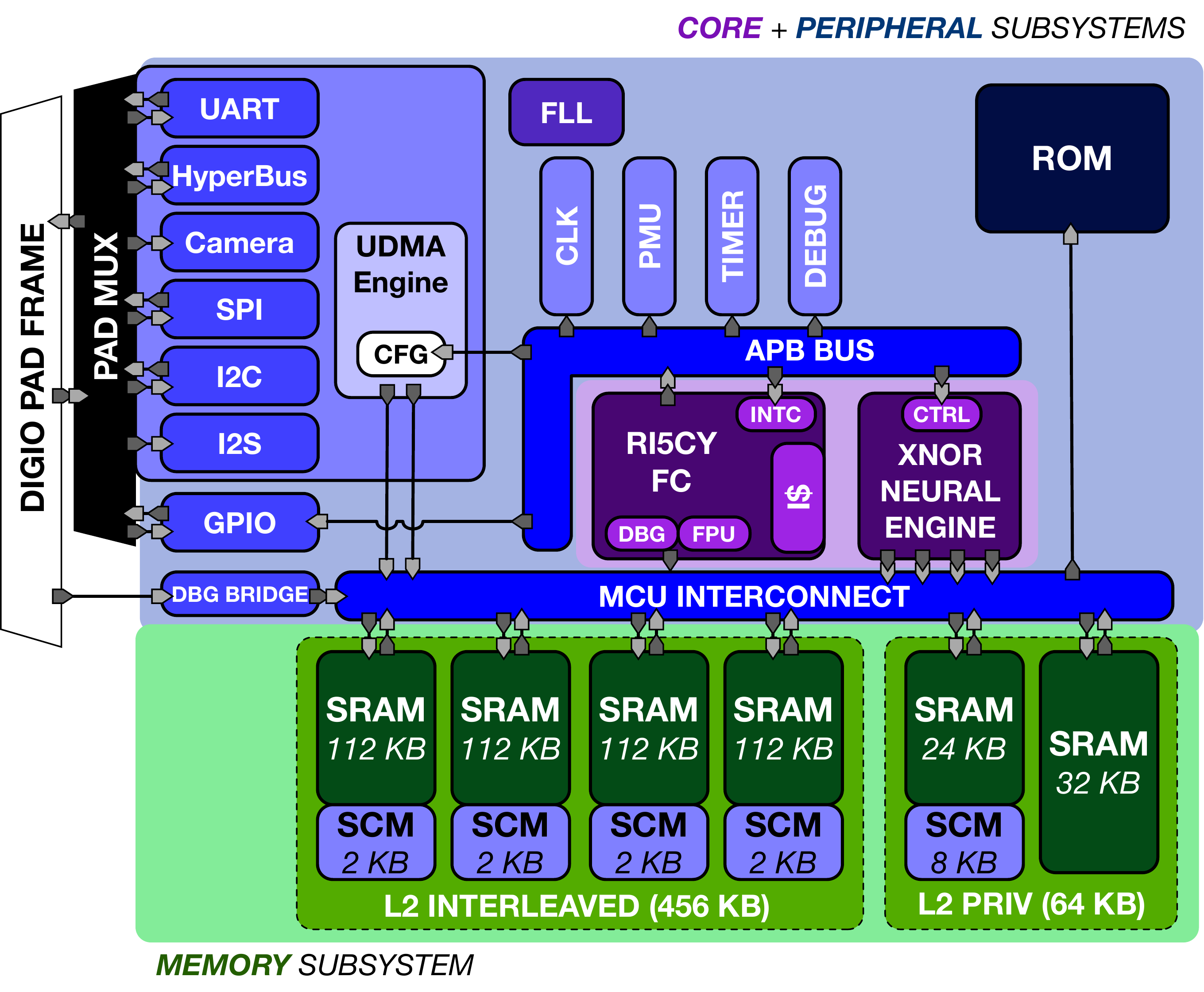}}
\caption{Quentin SoC Architecture. The \textit{core} subsystem is highlighted in violet, the \textit{peripheral} one in blue, and the \textit{memory} one in green.}
\label{fig:archi}
\end{figure}

This section introduces the system architecture of Quentin SoC, focusing on the micro-architecture of the binarized neural network accelerator (XNE) and on the heterogeneous system architecture and its implementation strategy enabling the power/performance/precision tunable capabilities of the system. The architecture of Quentin is reported in \figurename \ref{fig:archi}.

\subsection{System Architecture}

The system in exam consists of an advanced microcontroller based on the open-source PULPissimo system\footnote{https://github.com/pulp-platform/pulpissimo}, part of the Parallel Ultra-Low-Power (PULP) platform\footnote{https://pulp-platform.org}. The SoC is built around a RISC-V processor (RI5CY)\cite{RISCY} optimized for energy-efficient digital signal processing. The core's pipeline features 4 stages, floating-point and it is fully compliant with the RV32IMFC ISA \cite{Waterman14therisc-v}. On top of the standard RISC-V ISA the processor features digital signal processing extensions targeting energy-efficient near-sensor data analytic. These extensions include hardware loops, the automatic increment of pointers accesses, bit manipulation instructions, fixed-point and packed single-instruction-multiple-data (SIMD) operations, and unaligned memory accesses.

The system features a full set of peripherals which include Quad-SPI (QSPI) supporting up to two external devices, I2C, 2x I2S, a parallel camera interface, UART, GPIOs, JTAG, and a DDR HyperBus interface to connect off-chip up to 64 Mbytes of external Dynamic RAM (DRAM) or FLASH memory, and a small ROM used to store the boot-code. An I/O DMA ($\mu$DMA\cite{UDMA}) autonomously manages data transfers through peripherals to minimize the workload of the processor. To improve the efficiency of the system and the flexibility of transfer from/to the peripherals each peripheral has a dedicated clock domain. Two Frequency Locked Loops (FLLs) adjust the frequency of the peripheral subsystem and core subsystem. Moreover, peripherals are equipped with clock dividers that allow fine-tuning the frequency of according to the desired bandwidth. 
This architecture allows to tune the performance of computation and IO transfers, minimizing the system-level power consumption for the desired performance target.

\subsection{Hybrid Memory Architecture}

The L2 memory of the proposed SoC, shown in the bottom part of Fig. \ref{fig:chip} consists of a heterogeneous memory architecture designed to operate on a wide voltage range and to optimize access to the different regions of the memory depending on their purpose. From an architectural point of view, the memory is composed of two regions. The first one is a 64 kB private memory that can be used by Fabric Controller (FC) for storing its program, the stack, and other private data. This portion of the memory, connected to the interconnect through two ports (e.g. one for instructions and one for data), is typically not shared with other initiators, hence it does not incur any kind of conflicts guaranteeing full bandwidth. The second portion, called \textit{L2 interleaved} memory (Fig. \ref{fig:chip}), is composed of four 114 KB banks that can be accessed in parallel by the masters (i.e., $\mu$DMA\cite{UDMA}, instruction, and data port) while minimizing the banking conflict probability thanks to the interleaved addressing scheme implemented by the interconnect. From a performance viewpoint, this memory organization enables transparent sharing of the L2, increasing by 4x the system memory bandwidth compared to the traditional single-port memory architecture typical of AHB-based MCUs \cite{_stmicroelectronics-1}, without the usage of power-hungry dual-port memories.

Both memory regions described above are heterogeneous also from the memory technology point of view, being implemented as a hybrid mix of SRAM and standard-cell based memory cuts (SCM). The SCMs are based on the architecture described in \cite{SCM}. Each of the interleaved banks has 112 kB of SRAM and 2kB of SCM, while the private banks have 8 KB of SCM as shown in Figure ~\ref{fig:archi} while the rest is implemented as SRAM. The SCM portion of the private bank is implemented as a 3-read 2-write ports register file: two of the read ports and one of the write ports are dedicated to the data and instructions interfaces of the RISC-V core while one read and one write ports are used by the interconnect arbiter for any other master node of the system. Despite the intrinsic flexibility of synthesizable IPs that make them more suitable to implement multi-port cuts, one of the main advantages of latch-based memories is the capability, empirically proven in this work, to operate reliably in a much wider supply voltage range than SRAMs. Moreover, they feature significantly smaller read and write energy with respect to traditional SRAMs, up to 4x depending on the configuration (i.e. leakage dominated vs. dynamic dominated) \cite{SCM}. On the other hand, they pay a significant area overhead with respect to SRAMs, that makes them suitable only for implementing very small memory regions, usually in the orders of few kB \cite{SCM}. 

\begin{figure}[t]
\centerline{\includegraphics[width=0.95\linewidth]{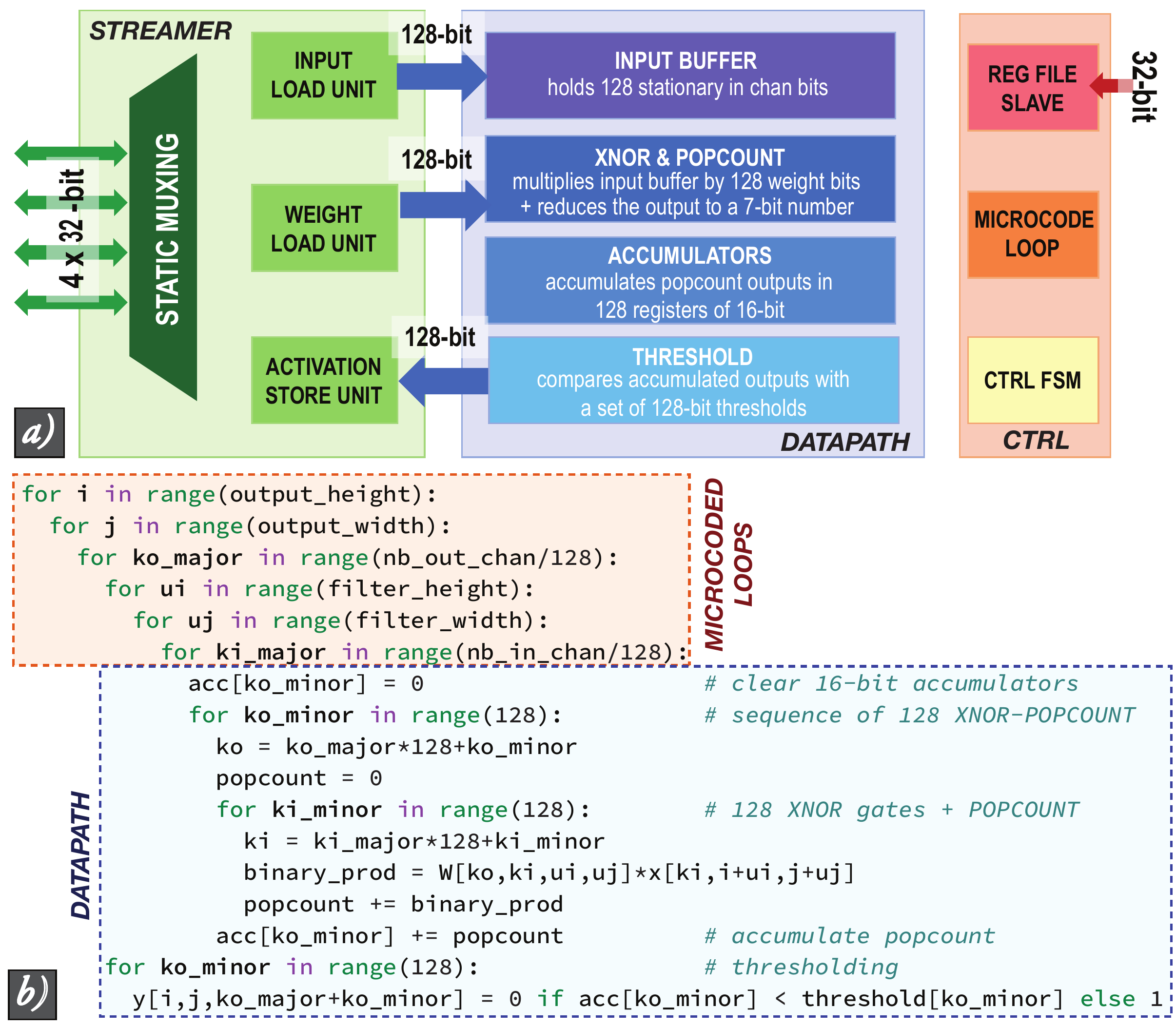}}
\caption{\textit{{a}) XNE internal architecture, showing the \textit{streamer} (green shades), \textit{control} (orange) and \textit{datapath} (blue) submodules; \textit{{b})} BNN layer execution pseudo-code highlighting microcoded loops (orange) and datapath execution (blue).}}
\label{fig:xne}
\end{figure}

\begin{figure}[t]
\centerline{\includegraphics[width=0.99\linewidth]{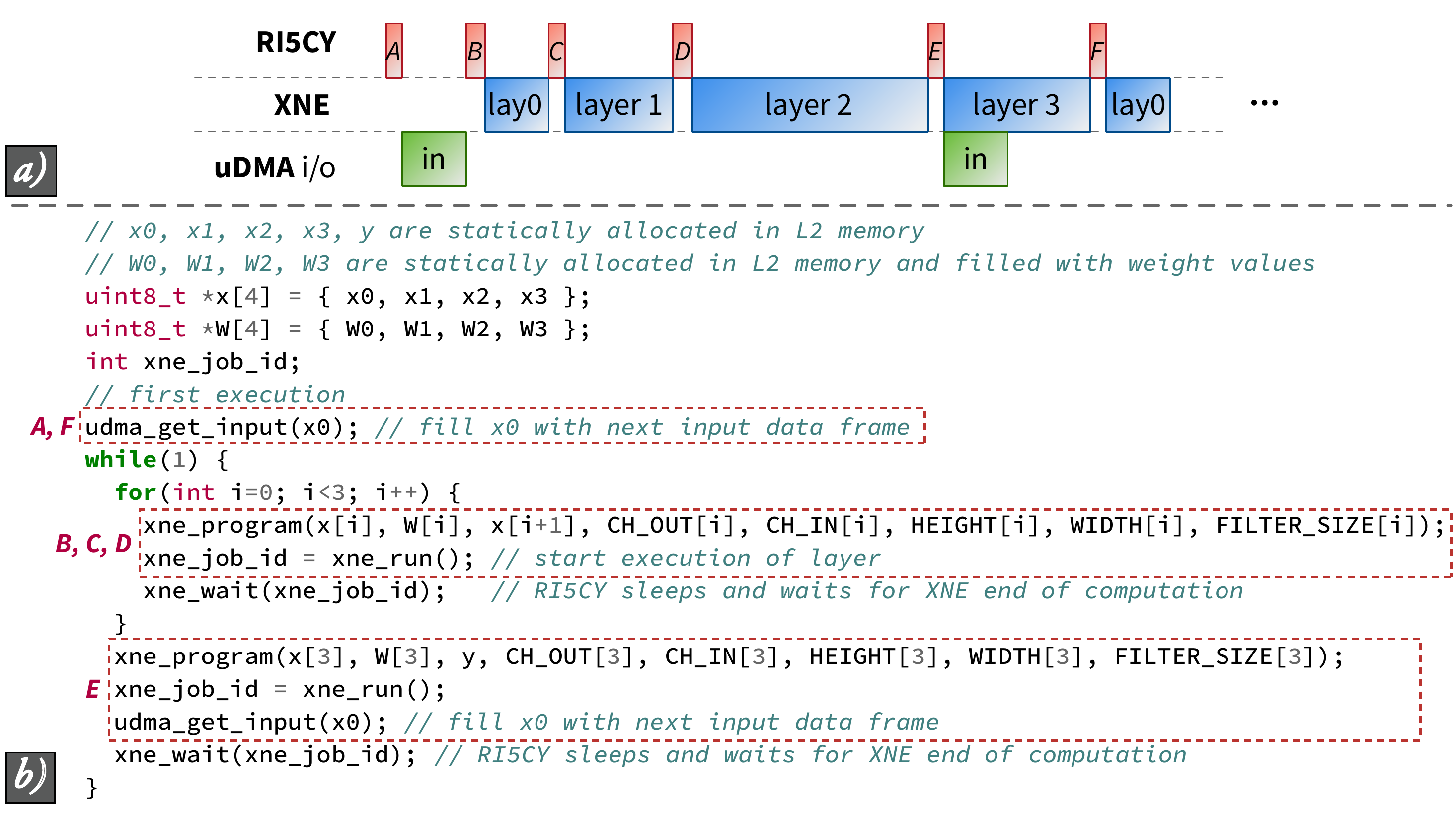}}
\caption{\textit{{a}) Execution profile for an example 4-layer fully on-chip BNN; \textit{{b})} ANSI C runtime code executed on RI5CY for the same 4-layer BNN.}}
\label{fig:sw}
\end{figure}

\subsection{XNOR Neural Engine and BNN Execution Model}

To execute with high performance and energy efficiency binary neural networks, the Quentin SoC also contains a dedicated hardware accelerator called XNOR Neural Engine (XNE)~\cite{ContiXNORNeuralEngine2018}.
The XNE is connected as a master to the interleaved L2 memory.
It has four ports, for an overall memory bandwidth of 128 bits per cycle.
\textcolor{black}{All the configuration registers are memory-mapped and accessible by the core. The XNE can execute both convolutional and fully connected layers, autonomously from the core, once all data reside in L2.}

Figure~\ref{fig:xne}a schematizes the internal architecture of the XNE.
It is divided in a \textit{control} submodule responsible of receiving jobs from the core; a \textit{streamer} submodule translating internal data streams into actual memory transfers on the memory interconnect towards L2; and a \textit{datapath} that performs binary matrix-vector products.
The controller includes a memory-mapped slave interface to a configuration register file, a controller finite-state machine and a small microcoded loop that is used to implement the following BNN layer execution pattern.

An \textsc{input buffer} is loaded with a stationary set of 128 input feature bits, with each bit representing a different input channel (0 represents a '-1' value, 1 a '+1' value).
The stationary input is multiplied with 128 bits of weights that are dynamically streamed in each cycle, using 128 parallel XNOR gates.
The \textsc{xnor} gates are followed by a 128-way reduction tree that performs a \textsc{popcount} operation.
Overall, the \textsc{xnor~\&~popcount} unit performs a full 128x128 binary matrix-vector product in 128 cycles, which is used to implement the innermost loops of a convolutional or linear BNN layer.
To implement the outer loops, popcount outputs are accumulated in a set of 128 registers (one per output channel) of 16-bit each.
After the accumulation is completed, the accumulated values are activated and binarized by comparing them with a set of 8-bit activation thresholds that are streamed in from memory and left-shifted by a configurable amount to be comparable with the 16-bit accumulators.
The execution is iterated as specified by the microcoded loop to implement a full BNN layer; if the granularity of the layer is smaller than 128 input or output channels, the datapath can be configured accordingly.
Figure~\ref{fig:xne}b describes the full execution schedule as pseudo-Python code; we refer to Conti~et~al.~\cite{ContiXNORNeuralEngine2018} for further detail.

Since the XNE operates at the granularity of a single BNN layer, the execution of a full network relies also on the operation of two other modules in the SoC: the RI5CY core, operating as a lightweight controller; and the UDMA engine, which is used to load inputs from I/O.
While the Quentin SoC is designed with the capability to access an external IoT DRAM if necessary, in this work we focus on the execution of relatively small, fully on-chip BNNs, which can be run within an ultra-low power budget by means of aggressive voltage scaling and access I/O exclusively to fetch input frames.
Figure~\ref{fig:sw} shows the execution profile of an example four-layer BNN, along with the C runtime code that is used to run it.
Runtime API calls wrap the memory-mapped control interfaces that both the UDMA and the XNE expose to configure them; therefore control is realized with fully compliant ANSI C code using regular load/store operations and requires no extension to the RISC-V ISA.
In the runtime code, \texttt{udma\_get\_input} API calls are synchronous and \texttt{xne\_run} asynchronous, with an explicit \texttt{xne\_wait} bringing RI5CY to sleep.

\subsection{Chip Implementation}

\begin{figure}[t!]
\centerline{\includegraphics[width=0.94\linewidth]{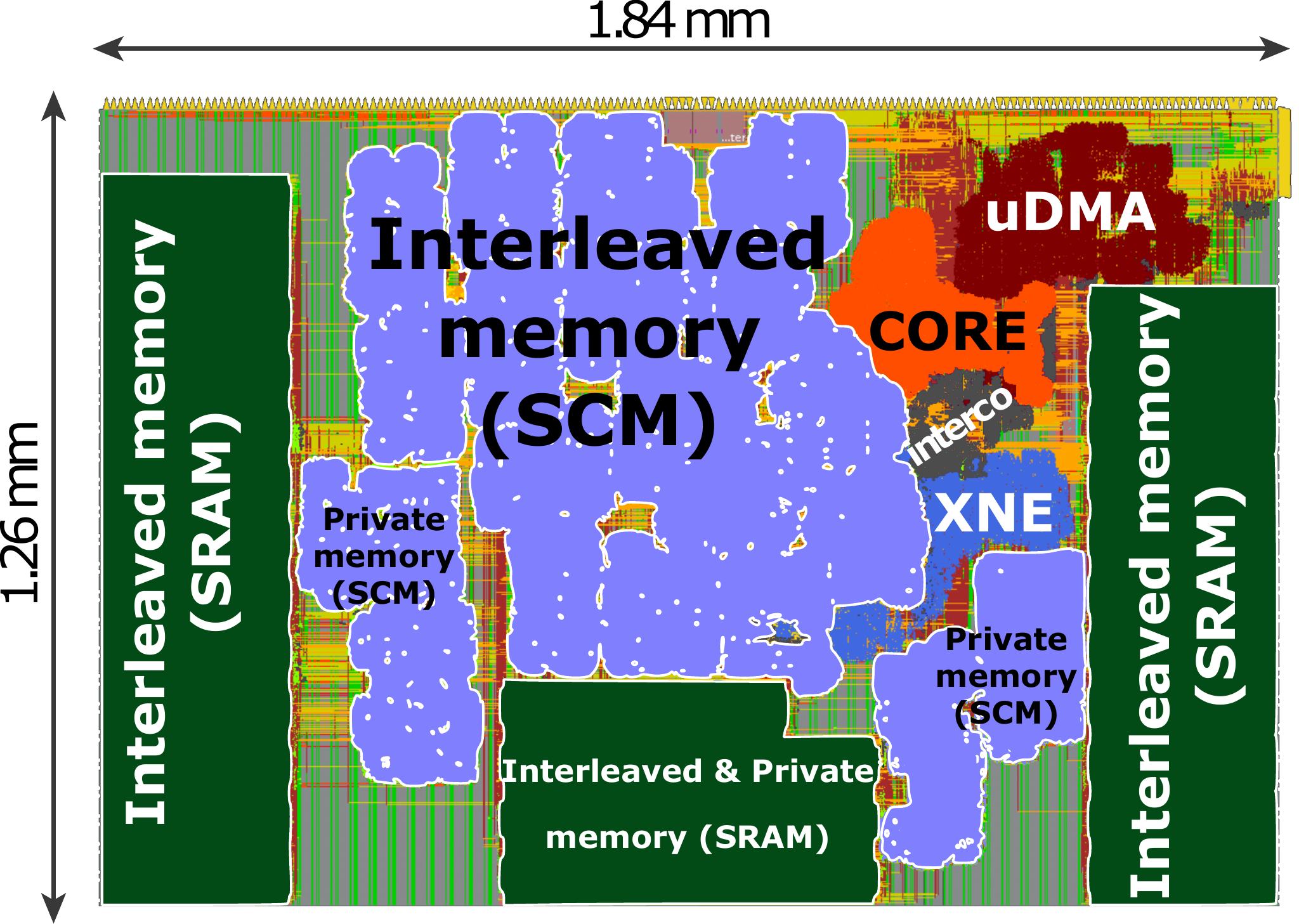}}
\caption{Quentin SoC floorplan.}
\label{fig:chip}
\end{figure}

\begin{table}[t!]
\caption{Quentin SoC features.}
\centering
\begin{tabular}{cc}
    \hline
	Technology                                & CMOS 22nm FD-SOI      \\
	Chip Area                                 & 2.3mm$^2$             \\
	Memory Transistors                        & 520 kbytes            \\
	Equivalent Gates (NAND2)                  & 1.8 Mgates            \\
	Voltage Range                             & 0.42 V -- 0.8 V       \\
	Body Bias Range                           & 0.00 V -- 1.4 V       \\
	Frequency Range                           & 32 kHz -- 670 MHz     \\
	Frequency Range (with FBB)                & 32 kHz -- 938 MHz     \\
	Power Range                               & 300 $\mu$W -- 10.4 mW \\
	Power Range (with FBB)                    & 300 $\mu$W -- 66.2 mW \\
	\hline
\end{tabular}
\label{tab:quentin_features}
\end{table}

\begin{figure}[t!]
\centerline{\includegraphics[width=0.95\linewidth]{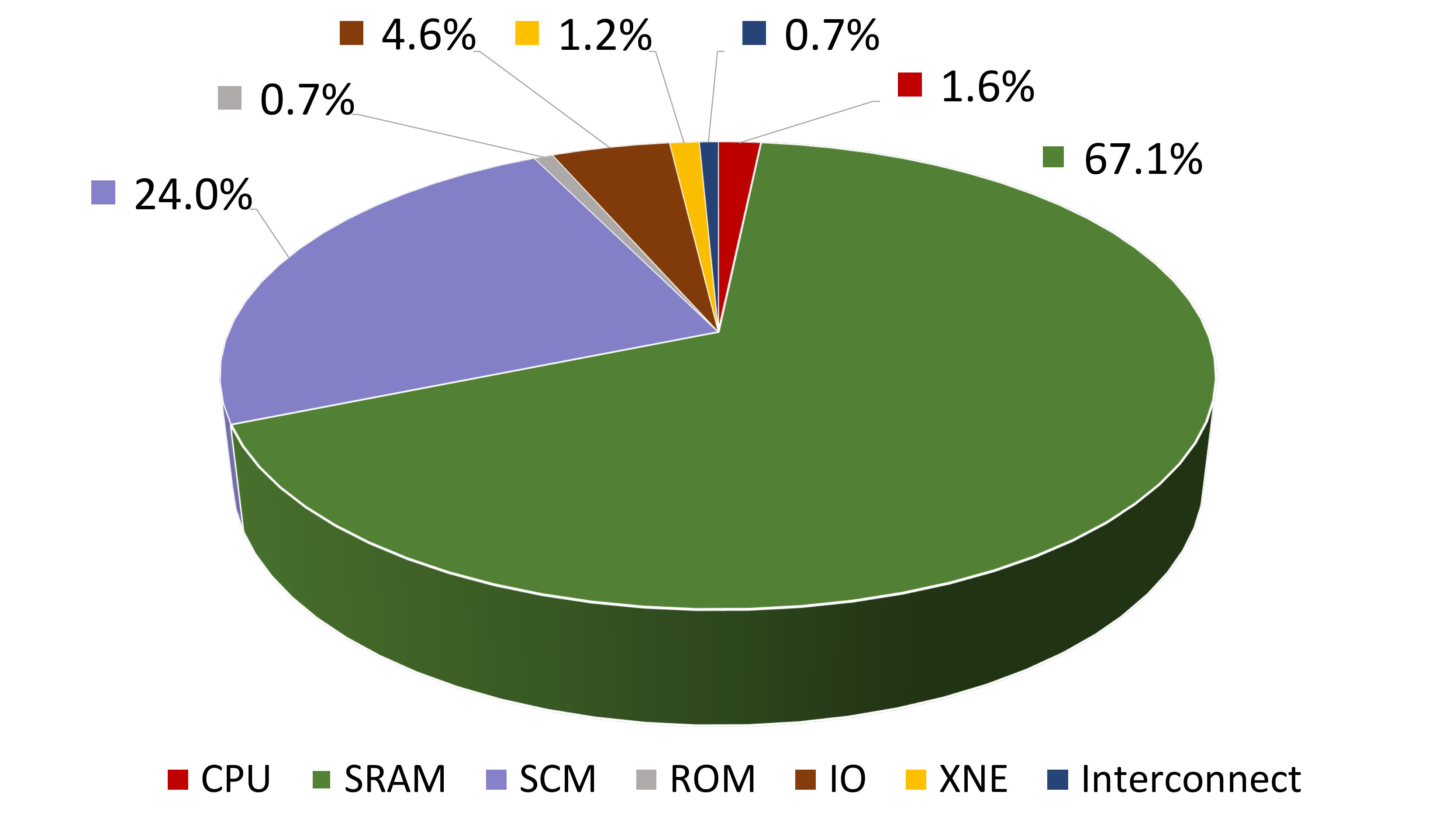}}
\caption{Quentin SoC area breakdown.}
\label{fig:piechart}
\end{figure}

\begin{figure}[b]
\centerline{\includegraphics[width=0.85\linewidth]{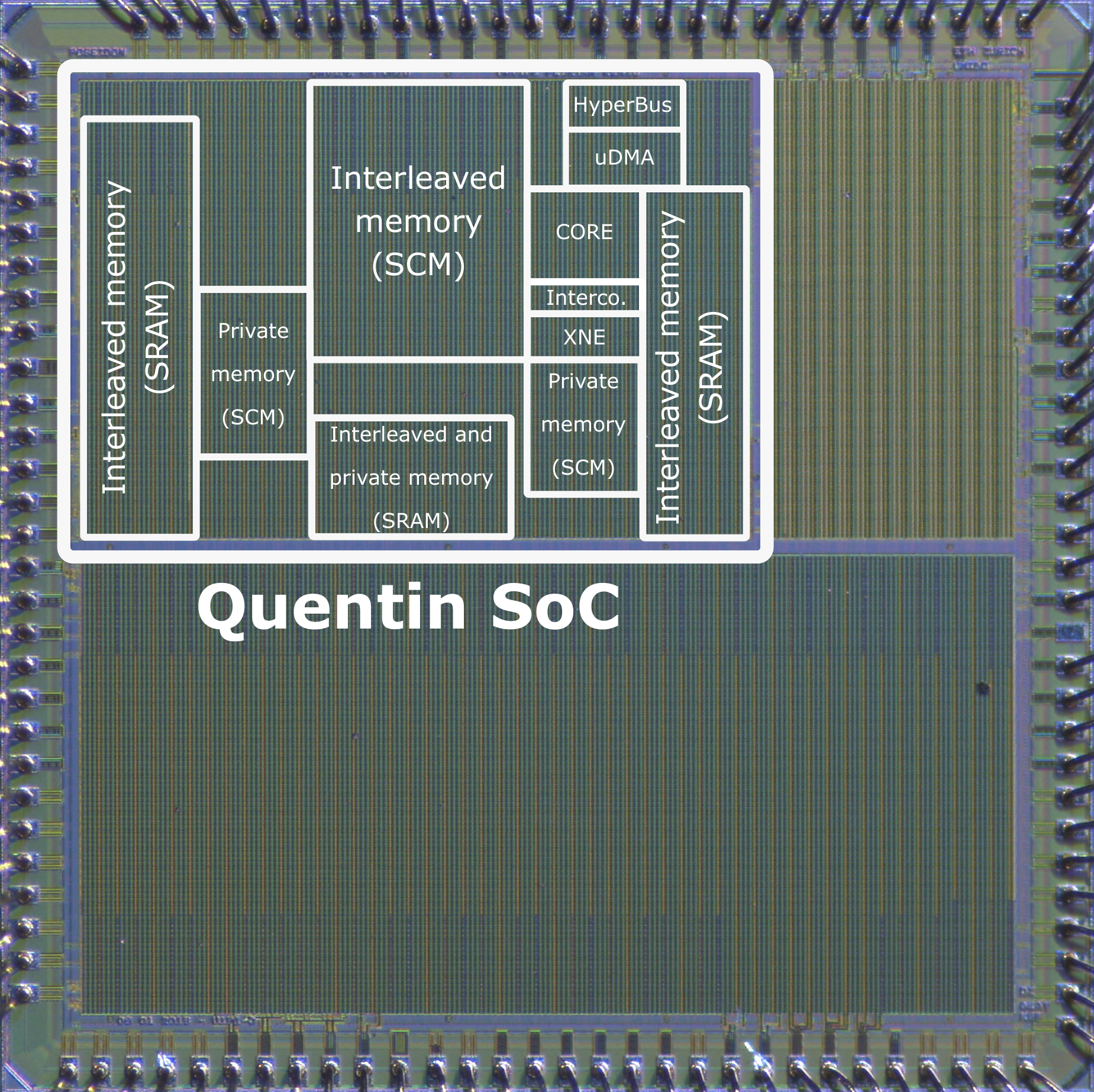}}
\caption{Quentin SoC chip micrograph.}
\label{fig:chip_micro}
\end{figure}

Figure \ref{fig:chip} shows the floorplan of the Quentin SoC, while Table \ref{tab:quentin_features} summarizes its main features. The SoC was implemented in 22nm FD-SOI technology using a flip-well (LVT) standard cell library.
The design was synthesized with Synopsys Design Compiler 2016.12, while Place $\&$ Route was performed with Cadence Innovus 16.10. \figurename~\ref{fig:chip_micro} shows a micrograph of the Quentin SoC\footnote{The area of the micrograph that is not annotated contains independent designs fabricated on the same chip}

\begin{table}[t!]
\caption{Quentin Area breakdown in mm\textsuperscript{2}}
\centering
\begin{tabular}{cc}
    \hline
	CPU subsystem                         & 0.020     \\
	SRAM (504kB)                          & 0.817     \\
	SCM  (16kB)                           & 0.292     \\
	ROM                                   & 0.009     \\
	I/O subsystem                         & 0.056     \\
	XNE                                   & 0.014     \\
	Interconnect                          & 0.009     \\
	\hline
\end{tabular}
\label{tab:quentin_area}
\end{table}

\begin{figure}[t]
\centerline{\includegraphics[width=0.95\linewidth]{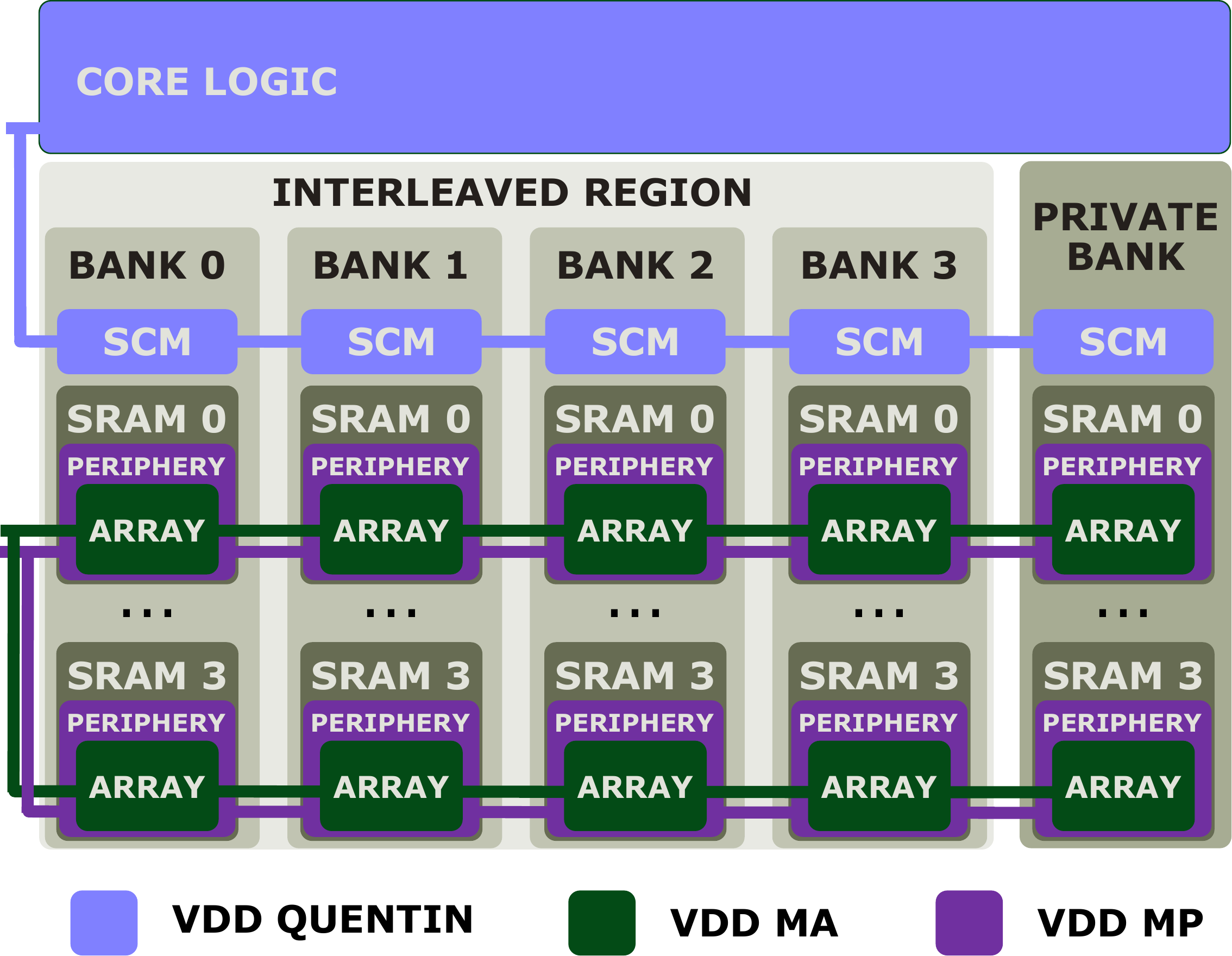}}
\caption{Quentin SoC Power Domains.}
\label{fig:power_domains}
\end{figure}

The floorplan area of the SoC is 2.31 mm\textsuperscript{2} and its effective area is 1.22 mm\textsuperscript{2} (6154KGE). Its main modules are highlighted in Figure \ref{fig:chip}. The two largest components of the SoC are the SRAM banks of the L2 memory subsystem (i.e., 504 kB), and by the 16 kB of SCM banks. \textcolor{black}{Although the latch based implementation features approximately a 10X area overhead compared to approaches based exclusively on SRAMs (Table \ref{tab:quentin_area}, \figurename \ref{fig:piechart}), it allows major energy savings \cite{SCM}, and it enables more flexible power management strategies that can be played at the system level. For example, SRAMs and SMCs can be independently power-gated. Additionally, on SCMs, it is possible to scale the operating voltage more aggressively than for SRAM. Our tests reported no errors when the supply voltage on the SCMs is scaled down to 0.42V; contrarily, errors on SRAMs become appreciable already at 0.575V (section \ref{sec:results}), limiting the voltage scaling capability of the system.}


To exploit both the energy advantage of SCMs and area density advantage of SRAMs, and to enhance the power/performance/precision tuning capabilities of Quentin, the chip was implemented as a multi power-domain system. The SRAM cuts have separate power connections from the rest of the logic for both periphery and array, as shown in Figure \ref{fig:power_domains}. This configuration allows us to independently tune the supply voltage of logic circuits, memory arrays, and memory periphery. Moreover, it allows the system to operate in an ultra-low-power, highly voltage-scaled mode using only the 16 KB SCM memories, and to shut down the SRAM via an off-chip power switch.


\section{BNN Error Resilience Analysis}
\label{sec:resilience_analysis}

As argued in Section~\ref{sec:intro}, BNNs have been shown to be partially resilient to high error rates.
For example, Yang~et~al.~\cite{Yang2018} use a statistical model to quantify the accuracy drop of a BNN in an application-specific architecture, reporting $\sim$5\% of accuracy loss with respect to the nominal accuracy under a bit error rate of $10^{-4}$.
\textcolor{black}{In this section, we evaluate the final classification accuracy of different pre-trained BNN topologies under multiple SRAM BER conditions. The goal of our analysis is to exploit the BNNs error resilience to enable major energy efficiency on SoC architectures featuring heterogeneous memory subsystems. Our results are silicon-proven on the Quentin chip. We performed our analysis on the CIFAR-10 classification data set.} 

\textcolor{black}{The BER reported in our experiments refers to data being fetched from the SRAM. In Quentin, the source where XNE fetches and stores binary data (i.e. weights, activations and partial results of internal BNN layers) is not fixed at design time.
As described in Section~\ref{sec:archi}, this data can be resident in either the interleaved SCM or SRAM memories;
the XNE accelerator always holds partial sums inside its accumulation buffer; only fully binarized outputs are stored back to the shared memory.} 

\textcolor{black}{In this scenario, we identified three potential sources of errors affecting the final BNN classification accuracy:  i) weights reading ii) input features reading iii) activations storage.
In our experiments, the XNE data-flow partial results are not affected by errors, as they are held inside the local buffer of the accelerator.
Additionally, output activations are binarized by comparing the final accumulation value $y$ with a safe 8-bit threshold value $\tau$, which is stored in the error-free SCM memory. Input features, weights, and activations reside in the SRAM, potentially corrupted by errors.}

\begin{figure}[tb]
\centerline{\includegraphics[width=\linewidth]{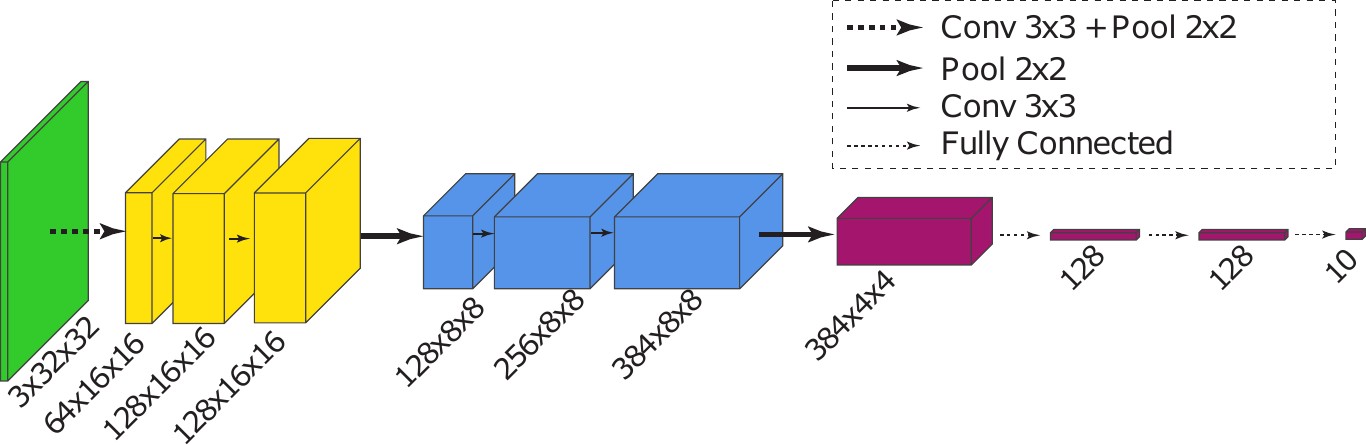}}
\caption{\textit{uVGG} BNN topology.}
\label{fig:our_bnn}
\end{figure}

To evaluate the accuracy loss when data are corrupted by a certain BER, we performed a set of simulations using the PyTorch 1.0.1 framework. We targeted a set of pre-trained networks on CIFAR-10 based on Hubara's implementation\footnote{https://github.com/itayhubara/BinaryNet.pytorch}.
We added uniformly distributed errors to input, weights, and activations of all layers of the network according to the target BER values to evaluate.
We tested the noisy BNNs on the test set of the CIFAR-10 data-set.
The networks have not been re-trained to compensate for the additional noise injected during the simulation.
This exploration ultimately allows evaluating the overall effect, in terms of final classification accuracy, of an aggressive voltage scaling performed on the SRAM. In other words, the SRAM supply voltage change can be performed dynamically, depending on the tolerable quality-of-results drop of a target application. In our experiments, we also explored the case where errors were occurring with recurring patterns. In this scenario, we did not observe any significant difference in the final classification accuracy with respect to the case where errors were uniformly distributed.

\begin{figure}[tb]
\vspace{-5mm}
\centerline{\includegraphics[width=\linewidth]{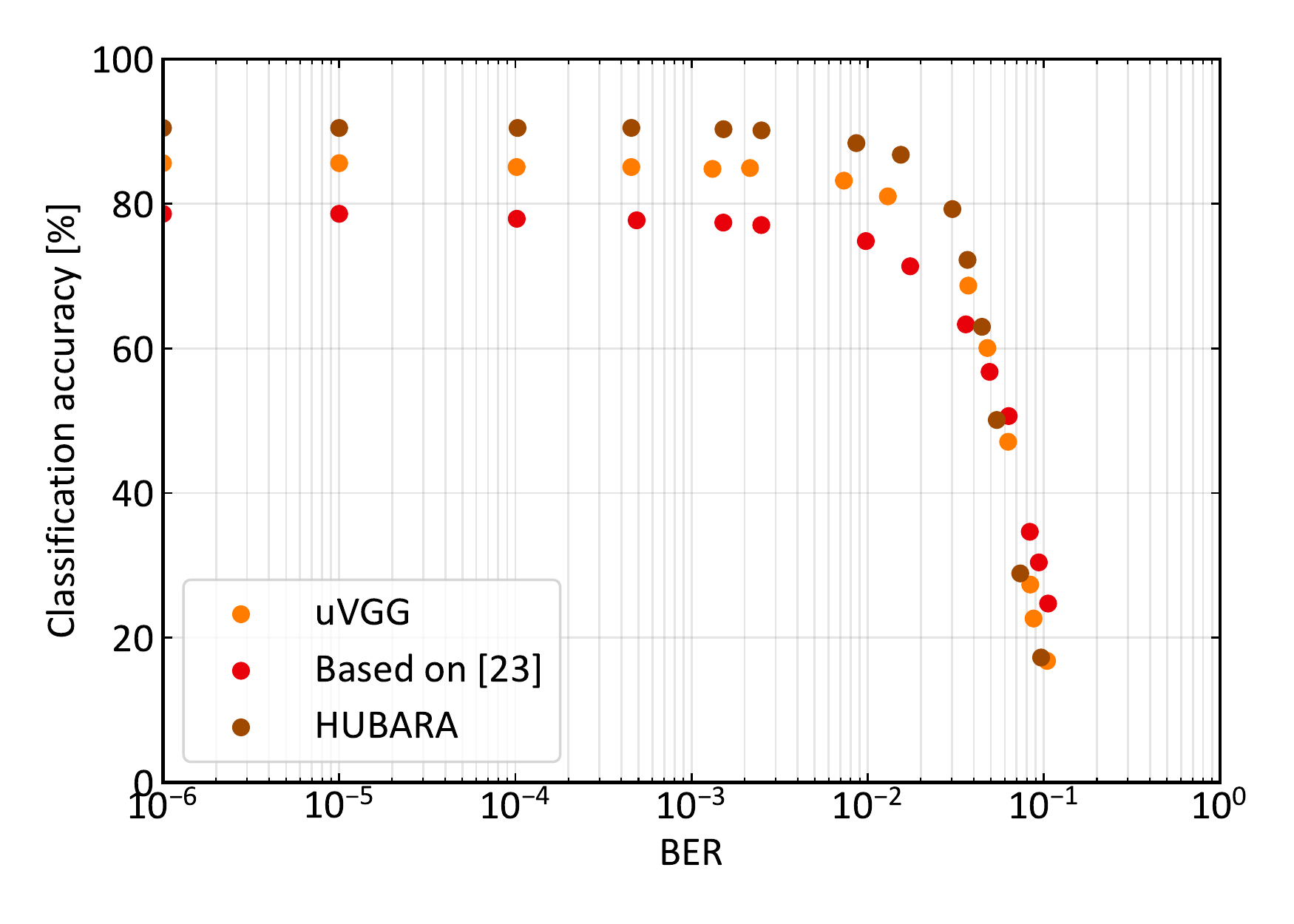}}
\caption{BNN resilience versus bit-error rate.}
\label{fig:accuracy_drop}
\end{figure}

\figurename~\ref{fig:accuracy_drop} shows the results in terms of classification accuracy versus the BER.
The classification accuracy of each network is reported as an average over 100 randomized experiments over the CIFAR-10 test set; the standard deviation of the results over this sample is always less than $1\%$ of the reported value.
We report results on Hubara's topology, as well as on a \textcolor{red}{network inspired by the one proposed by Yang~et~al.}
We also report results on our topology, similar to Hubara's~\cite{HubaraQuantizedNeuralNetworks2016} but fit to be deployable on the Quentin SoC.
Figure~\ref{fig:our_bnn} shows the topology of this latter network, which we called \textit{micro}-VGG (uVGG).
Table~\ref{tab:bnn_param} reports the salient characteristics of these networks.
For what concerns the network proposed by Yang~et~al.~\cite{Yang2018}, we \textcolor{red}{were not able to reproduce the exact training setup.
Our PyTorch implementation, on which our experimental results are based, achieves significantly lower accuracy results than what Yang~et~al.~\cite{Yang2018} report (78.6\% instead of 85\%)}.

\begingroup
\renewcommand*{\thefootnote}{\alph{footnote}}
\begin{table}[t]
\centering
  \begin{threeparttable}
    \begin{savenotes}
    \begin{tabulary}{\textwidth}{C C C}
      \toprule
      \textbf{BNN topology}      & \textbf{Nominal accuracy }      & \textbf{Mem. footprint} \\
      \midrule 
      \textcolor{red}{Based on} Yang~et~al.~\cite{Yang2018}       & 78.6\%\footnotemark[1] & 319 kB  \\
      Hubara~et~al.~\cite{HubaraQuantizedNeuralNetworks2016},  & 90.9                   & 4545 kB \\
      \textit{uVGG}                & 85.6\%                 & 312 kB  \\
      \bottomrule
    \end{tabulary}
    \end{savenotes}
    \begin{tablenotes}
      \item \footnotemark[1]{} Including both activations and weights.
    \end{tablenotes}
  \end{threeparttable}
\caption{Parameters of BNNs used in the resilience experiment.}
\label{tab:bnn_param}
\end{table}
\endgroup

In the remainder of the paper, we focused on our proposed network, which can tolerate a BER of $10^{-4}$ with negligible accuracy drop (\textless4\textperthousand) and a BER of $10^{-3}$ with an accuracy drop of 7\textperthousand, while fitting perfectly in the SRAM of the Quentin SoC.

\section{Experimental Results}
\label{sec:results}

In this section, we describe the results of experiments that allowed us to correlate the SRAM supply voltage scaling to the classification accuracy of the uVGG BNN presented in Section~\ref{sec:resilience_analysis}.
As a first step, we evaluated the level of noise that could potentially corrupt the data stored in the SRAM by measuring how the Bit Error Rate (BER) correlates with the memory array and peripheral voltage supplies.
As a second experiment, we measured the current drained by each power domain of the SoC to extract the power consumption. The power contribution reported in this section refers to the independent power rails described in \figurename \ref{fig:power_domains}.
Finally, we computed the energy efficiency of the SoC and evaluated the power saving when the supply voltage of the SRAM is scaled and the quality-of-results (i.e. top1 network accuracy) is degraded by less than 1\%. 

\subsection{Experimental setup}
\label{subsec:setup}

\begin{figure}[t]
\vspace{-5mm}
\centerline{\includegraphics[width=\linewidth]{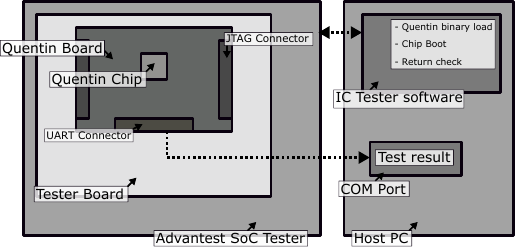}}
\caption{High level block diagram of the experimental setup.}
\label{fig:tester}
\end{figure}

All the measurements related to SRAM BER evaluation and power domains power consumption have been performed using an Advantest SoC hp93000 integrated circuit testing device. Supply voltages have been precisely regulated utilizing dedicated hp93000 power supply device channels. Power measurements have been performed using current measurement channels integrated into the hp93000 device and connected in series to the voltage supply channels. 

BER experimental data have been obtained by running a self-test C application on the RISCY microprocessor executing from the private SCM of the core, which is error-free in all operating points tested. We loaded the test program into the SCM through the SoC internal debug interface via a standard JTAG interface driven by the hp93000 digital channels. \figurename \ref{fig:tester} shows the block diagram of the experimental setup.




\subsection{Bit Error Rate analysis}

Measuring Bit Error Rates from outside, i.e. directly from the tester equipment,  requires a very large testing time. In our tests, we observed that the number of bits to observe to detect a single-bit memory failure can be in the order of $10^{9}$ or higher, e.g. for SRAM operating at nominal supply voltage conditions.
Additionally, to acquire relevant statistics on memory errors, tests have to be repeated many times. To reduce the number of pads dedicated to SoC debug subsystems, modern micro-controllers often employ serial debug interfaces connected to a shared bus. Therefore, accessing the memory locations through a serial JTAG debug interface designed for reliability rather than for high speed, could be a severe limitation for the execution of tests targeting BER measurement. \textcolor{black}{ In our tests, we estimated that a single BER measurement point is acquired in several 10th of minutes, assuming to test 448kB of memory for 1800 iteration, at a JTAG frequency of 1MHz, and repeating each measurement 10 times. To overcome the serial debug interface bottleneck, we designed an on-chip BER test application, which was executed by the microcontroller core. This allowed to reduce the time to test a single BER point by a factor of approximately 100X.}

To issue memory transactions to the SRAM, and observe errors on the bits, our self-test application runs directly on the RISCY core of the SoC, which operates at the highest reliable frequency for each condition.
Pseudo-random test patterns are generated by the core using a lightweight 32 bits Linear Feedback Shift Register (LFSR) implemented in C code. The test application sequentially covers the entire SRAM shared address space. Errors are counted by comparing, bit-wise, the data read at each memory location with the ground-truth value generated by the LFSR generator using the same initial seed. At each supply voltage point, the test is repeated in a loop to have a reliable measurement of the BER.
Note that this approach could generate artifacts in the error statistics when a memory location is filled in successive iterations with the same test vector; to avoid this problem, and to make our measurement more robust, the software LFSR uses a different seed to generate test data at each new iteration.

In our tests, we measured only the BER related to SRAM banks.
SCM, which is hosted by the same power domain as the circuit logic, was reserved for storing the core instructions of the self-test application and test results (i.e. the number of errors).
Note that the storage of the software instruction on an error-free memory space is mandatory for the application to be able to run.
In SoCs featuring single-power-domain memory subsystems (i.e. not having the possibility to store core instructions in a separate error-free memory), SRAM errors could affect also core instructions -- making aggressive voltage scaling infeasible, as a single corrupted bit on a core instruction could cause errors in the core control flow, making the entire SoC entering unpredictable states, and ultimately the system to fail.
For each operating point in our experiments, we performed 1800 on-chip test runs, writing 448kB at each iteration. 


\begin{figure}[tb]
\centerline{\includegraphics[width=\linewidth]{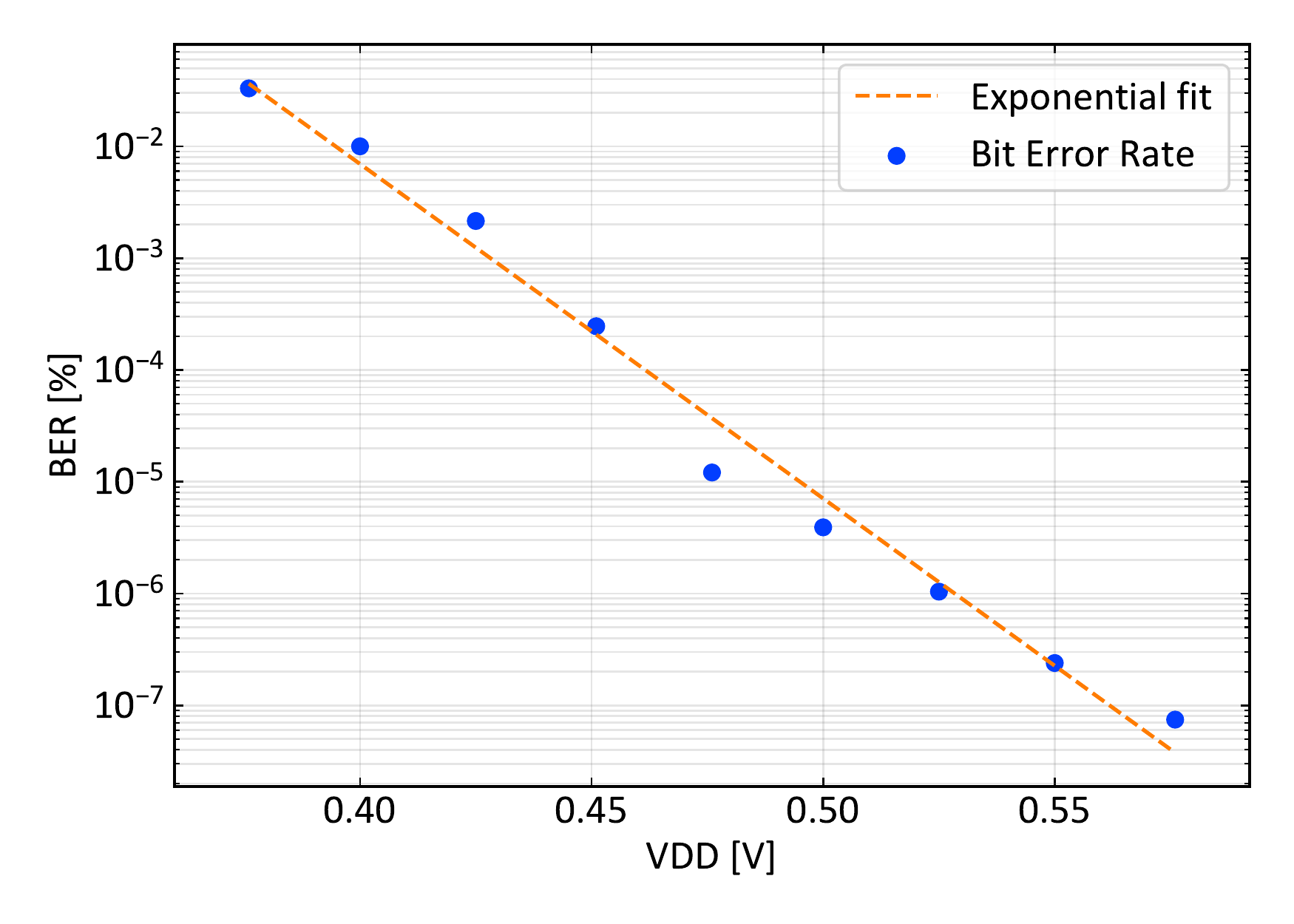}}
\caption{\textcolor{black}{Bit Error Rate.}}
\label{fig:ber_all}
\vspace{-4mm}
\end{figure}

\begin{table}[b]
    \centering
    \begin{tabular}{ccc}
        \toprule
        \textbf{OP mode} & $Vdd_{ma/mp/quentin}$ & \textbf{Freq.}\\
        \midrule
        Nominal & 0.8V & 565~MHz \\
        HEFF & 0.5V & 145~MHz\\
        ULP & 0.42V & 18~MHz \\
        \bottomrule
    \end{tabular}
    \caption{\textcolor{black}{Supply voltage range of Memory Array (MA), Memory Periphery (MP) and Quentin power domains at Nominal, High Efficiency (HEFF) and Ultra-Low Power (ULP) modes.}}
    \label{tab:op_conditions_ber}
\end{table}

\figurename~\ref{fig:ber_all} reports the BER at each SoC operating voltage.
By construction, our test could not observe more than $8*10^{8}$ bits. Therefore, the reciprocal of this value represents the lower bound of the on-chip test application, i.e. $1.25*10^{-9}$. The results of the BER analysis versus the supply voltage are reported in \figurename \ref{fig:ber_all}. When the supply voltage is higher than \SI{0.6}{\volt}, no BER is observable by our tests.

Below a supply voltage of \SI{0.6}{\volt}, as expected, we observed a BER increasing with the memory supply voltage decrease, reaching a BER of $10^{-2}$ at the lowest supply voltage point where the memory was still accessible. 
The BER measurements confirm that SRAM supply voltage can be scaled at the cost of a higher number of errors, noise-tolerant applications can be deployed on Quentin SoC and there is enough margin for trading off the amount of noise injected on the data and the potential energy efficiency gain deriving from the voltage scaling.



\begin{figure}[t]
\centerline{\includegraphics[width=1\linewidth]{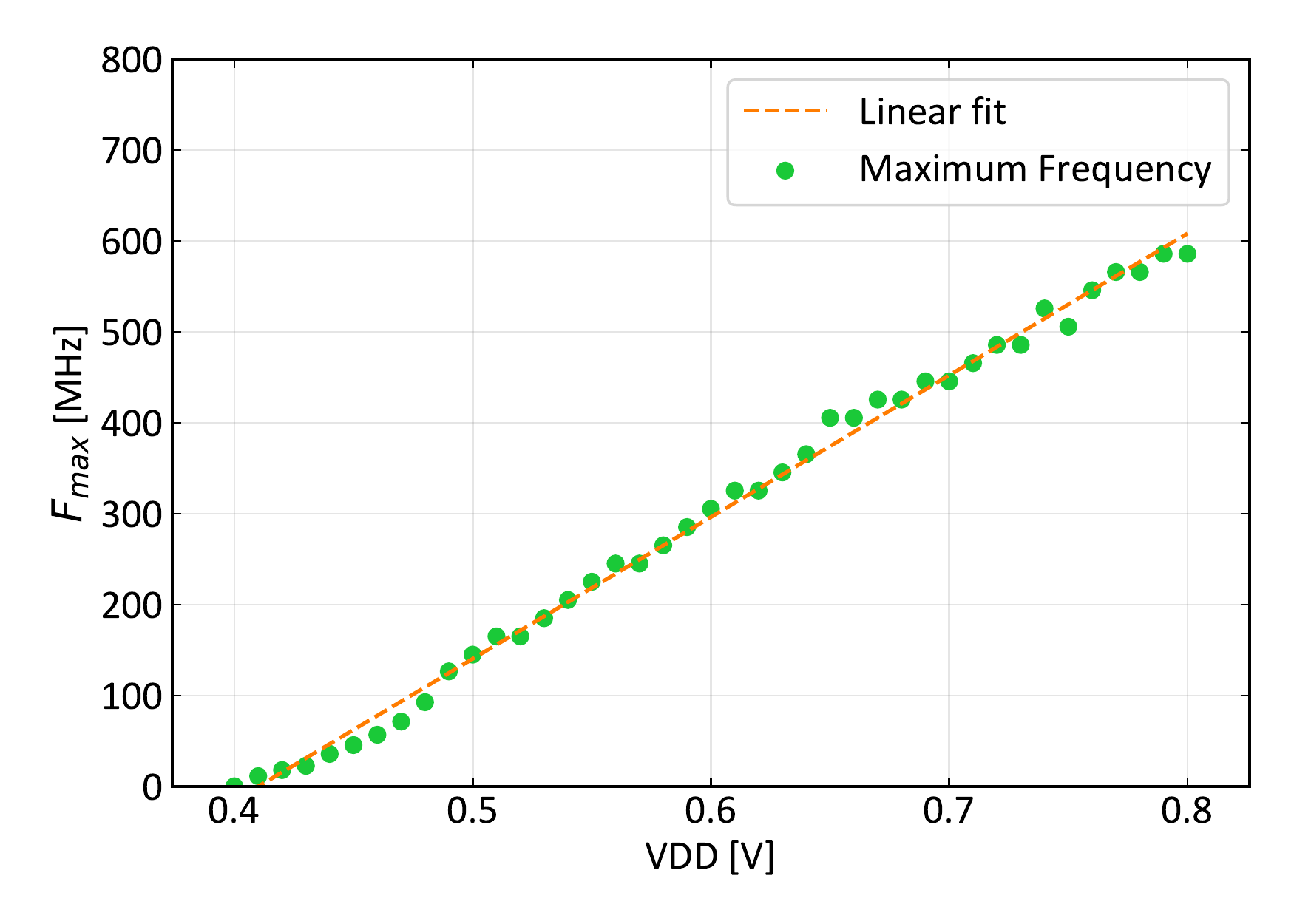}}
\caption{\textcolor{black}{SoC maximum operating frequency.}}
\label{fig:max_freq}
\end{figure}

\subsection{Power and energy consumption}

In this section, we discuss results related to the power and energy consumption of the SoC. These measurements, together with the evaluation of the maximum operating frequency, which is reported in \figurename \ref{fig:max_freq}, allow evaluating the overall energy efficiency of the system. The critical path of the system is in the paths going from the core to the memory system. Power measurements were performed during the execution of a test application on the Quentin SoC. To precisely control the supply voltages and clock frequencies of the SoC, therefore to measure the energy consumption of individual SoC power domains with enough accuracy, all the measurements were performed on the Advantest SoC IC tester mentioned in \ref{subsec:setup}.

\begin{table}[b]
\begin{tabular}{@{}cccccc@{}}
\toprule
 & \textbf{Weights} & \textbf{\begin{tabular}[c]{@{}c@{}}Activation \\ thresholds\end{tabular}} & \textbf{\begin{tabular}[c]{@{}c@{}}Input \\ features\end{tabular}} & \textbf{\begin{tabular}[c]{@{}c@{}}Output \\ features\end{tabular}} & \textbf{\begin{tabular}[c]{@{}c@{}}Instructions\\ and Stack\end{tabular}} \\ \midrule
\textbf{\begin{tabular}[c]{@{}c@{}}SRAM \\ Exec.\end{tabular}} & {\color[HTML]{9A0000} \textit{SRAM}} & {\color[HTML]{036400} \textit{SCM}} & {\color[HTML]{9A0000} \textit{SRAM}} & {\color[HTML]{9A0000} \textit{SRAM}} & {\color[HTML]{036400} \textit{SCM}} \\
\textbf{\begin{tabular}[c]{@{}c@{}}SCM\\ Exec.\end{tabular}} & {\color[HTML]{036400} \textit{SCM}} & {\color[HTML]{036400} \textit{SCM}} & {\color[HTML]{036400} \textit{SCM}} & {\color[HTML]{036400} \textit{SCM}} & {\color[HTML]{036400} \textit{SCM}} \\ \bottomrule
\end{tabular}
\caption{Network parameters and core instruction memory storage}
\label{tab:memory_storage}
\end{table}

The application we used as a benchmark was designed to emulate realistic working conditions, using the XNE accelerator over a randomized pattern of bits. By using a synthetic uniformly distributed and uncorrelated set of binary inputs, we maximized the switching activity both on the XNE input circuitry and the XNOR-based datapath, practically obtaining a worst-case power consumption.
The main application is executed by the core while the BNN layer output computation is accelerated by the XNE; while the accelerator is active, the core is in clock-gating and it wakens up at the end of each XNE job.
Instruction code was stored in the error-free SCM, supplied at the same voltage as the core. Data (i.e. binary weights, activation, and partial results) were stored in the error-prone SRAM supplied at the scaled voltage, except for critical 8-bit threshold data which is stored in the interleaved SCM. Table \ref{tab:memory_storage} reports more details on how data are stored in memory.

The measurements reported in this section refer to three relevant operating conditions of the SoC. The nominal operating point (namely Nominal) refers to a supply voltage of \SI{0.8}{\volt}. This is the operating point for which we performed the Static Timing Analysis (STA). The High Efficiency (HEFF) point is the operating condition at which the chip reported the highest energy efficiency. The Ultra-Low Power (ULP) point is the operating condition where the chip reported the minimum power consumption. Table~\ref{tab:op_conditions_ber} provides more details about those three points.

\begingroup
\begin{table*}[t]
\begin{tabular}{@{}lcccccccc@{}}
\toprule
\textbf{Name} & \textbf{Technology} & \textbf{\begin{tabular}[c]{@{}c@{}}Core area\\ {[}$mm^{2}${]}\end{tabular}} & \textbf{\begin{tabular}[c]{@{}c@{}}Power\\ {[}$mW${]}\end{tabular}} & \textbf{\begin{tabular}[c]{@{}c@{}}Energy eff.\\ {[}$TOP/s/W${]}\end{tabular}} & \textbf{\begin{tabular}[c]{@{}c@{}}On-chip memory\\ {[}$kByte${]}\end{tabular}} & \textbf{\begin{tabular}[c]{@{}c@{}}Peak perf.\\ {[}$GOP/s${]}\end{tabular}} & \textbf{Type} & \textbf{BNN} \\ \midrule
\textit{BRein}~\cite{Ando2018} & \begin{tabular}[c]{@{}c@{}}65nm\\ (Digital)\end{tabular} & 3.9 & \textcolor{red}{600} & 6 & - & \textcolor{red}{1380} & DNN ASIC & Configurable \\ \midrule
\textit{UNPU}~\cite{UNPU2019} & \begin{tabular}[c]{@{}c@{}}65nm\\ (Digital)\end{tabular} & 16 & 7.37 & 51 & 256 & 7372 & DNN ASIC & Configurable \\ \midrule
Bankman~et~al.~\cite{BankmanAlwaysOn8mJ86} & \begin{tabular}[c]{@{}c@{}}28nm\\ (Mixed signal)\end{tabular} & 4.84 & 0.094 & 772 & 329 & 72 & DNN ASIC & Fixed topology \\ \midrule
BinarEye~\cite{moons2018} & \begin{tabular}[c]{@{}c@{}}28nm\\ (Digital)\end{tabular} & 1.4 & 2.2 & 230 & 328 & 90 & DNN ASIC & Configurable \\ \midrule

Yin et al.~\cite{Liu2019} & \begin{tabular}[c]{@{}c@{}}28nm\\ (Digital)\end{tabular} &  4.8 & 3.4-20.8 & 765 & 224 & 3270 & DNN ASIC & Configurable \\ \midrule




This work (0.8V) &
  \multirow{4}{*}{\begin{tabular}[c]{@{}c@{}}22nm\\ (Digital)\end{tabular}} &
  \multirow{4}{*}{2.3} &
  21.6$^a$ &
  5.98$^a$/14$^b$ &
  \multirow{4}{*}{520} &
  129$^a$ &
  \multirow{4}{*}{\begin{tabular}[c]{@{}l@{}}Heterog. SoC\\ core + periph + \\ mem + BNN acc.\end{tabular}} &
  \multirow{4}{*}{SW defined} \\ \cmidrule(r){1-1} \cmidrule(lr){4-5} \cmidrule(lr){7-7}
\begin{tabular}[c]{@{}l@{}}This work (0.8V)\\ SW Impl.\end{tabular} &  &  & 16$^a$    & 0.276$^a$   &  & 3.73$^b$ &  &  \\ \cmidrule(r){1-1} \cmidrule(lr){4-5} \cmidrule(lr){7-7}
This work (0.5V)                                                    &  &  & 2.52$^a$  & 13$^a$/23.9$^b$ &  & 33$^a$   &  &  \\ \cmidrule(r){1-1} \cmidrule(lr){4-5} \cmidrule(lr){7-7}
This work (0.42V)                                                   &  &  & 0.674$^a$ & 6.2$^a$/14$^b$  &  & 4$^a$    &  &  \\ \bottomrule

\end{tabular}

   \vspace{1ex}
   {a - full SoC \\ b - core domain}
   
   \caption{Comparison of silicon-proven Application-Specific ICs for Binary Neural Networks.}
    \label{tab:comparison}
\end{table*}
\endgroup




\begin{figure}[t]
\centerline{\includegraphics[width=1.05\linewidth]{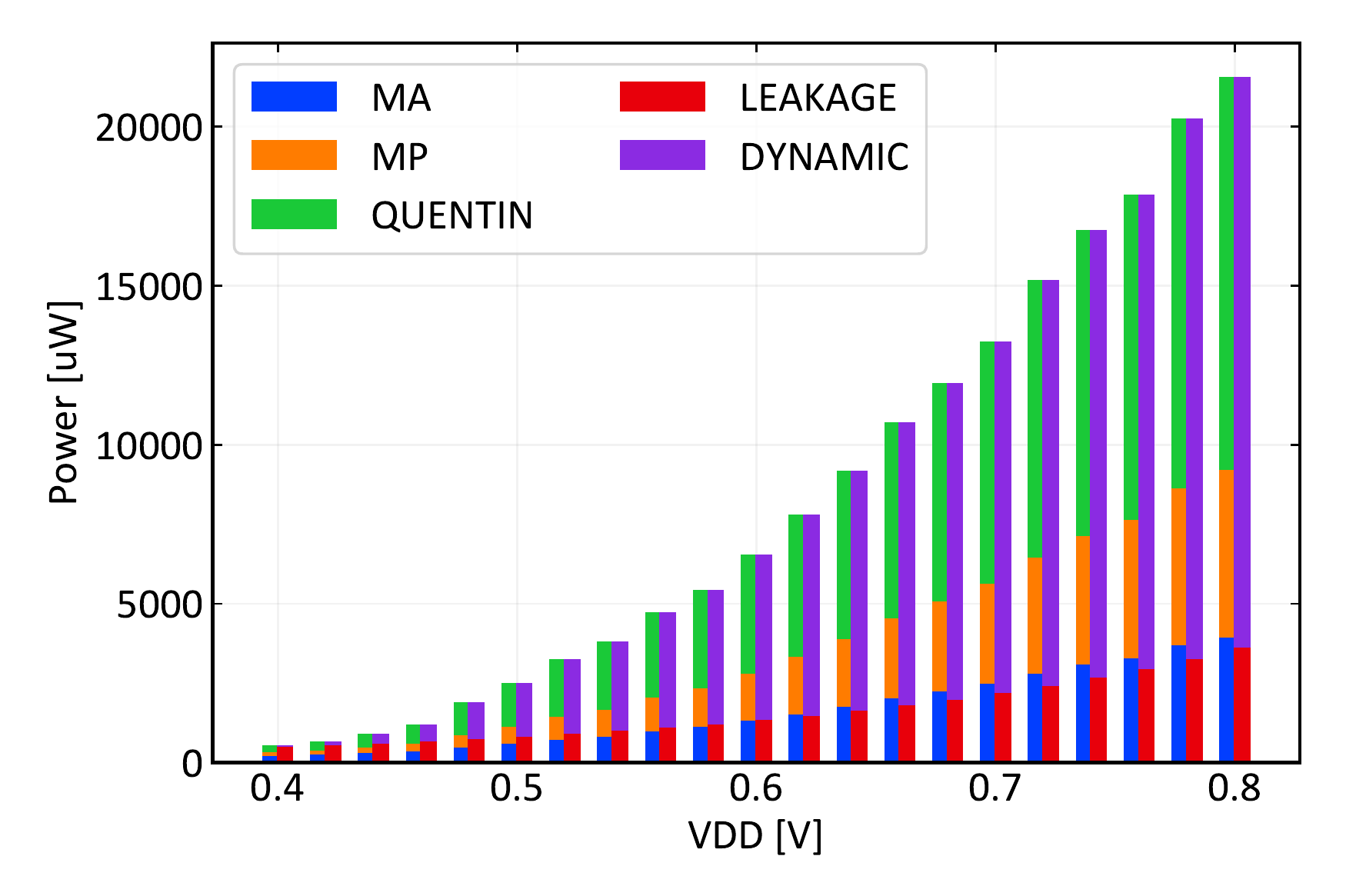}}
\caption{Breakdown of independent power contributions when operating from supply voltage scaled SRAM.}
\label{fig:power_analysis}
\end{figure}

\figurename \ref{fig:power_analysis} reports the power consumption breakdown of the Quentin SoC. We measured contributions from the three power domains by observing the total current drained from the power supply by each power domain. We performed all power measurements at three different frequencies (fmax, fmax/2, 10 MHz) and used a simple least squares model to detect static and dynamic power.
As shown by the plot in \figurename \ref{fig:power_analysis}, in ULP mode (i.e. at a supply voltage of \SI{0.42}{\volt}), leakage power dominates the dynamic power. In this operating point, the SoC reliably works at its lowest power consumption, \SI{674}{\micro \watt}, yet achieving a sustainable frequency of 18~MHz, and energy efficiency of 6.2~Tops/s/W, which is not lower than the one reported at the nominal operating condition. In ULP mode, the leakage power contributes to approximately 80\% of the total power consumption. In this operating point, Quentin can perform 15.4~Inference/s, reaching an energy efficiency of 22.8~Inference/s/mW.
When compared to the execution of the same workload on the embedded RI5CY core, hardware-accelerated execution achieves 21.6$\times$ better energy efficiency in the Nominal operating point (167 fJ/op using the XNE, 3.6 pJ/op performing it in software at VDD=0.8V): whereas the overall power consumption drops when using the core, the performance achieved is significantly lower (6.6 op/cycle)~\cite{garofalo2019pulpnn}. Note that the software baseline used as a comparison represents already a significant improvement (more than 10X) to the performance reported by leading microcontroller architecture (e.g. arm cortexM4).

\begin{figure}[t]
\centerline{\includegraphics[width=1.05\linewidth]{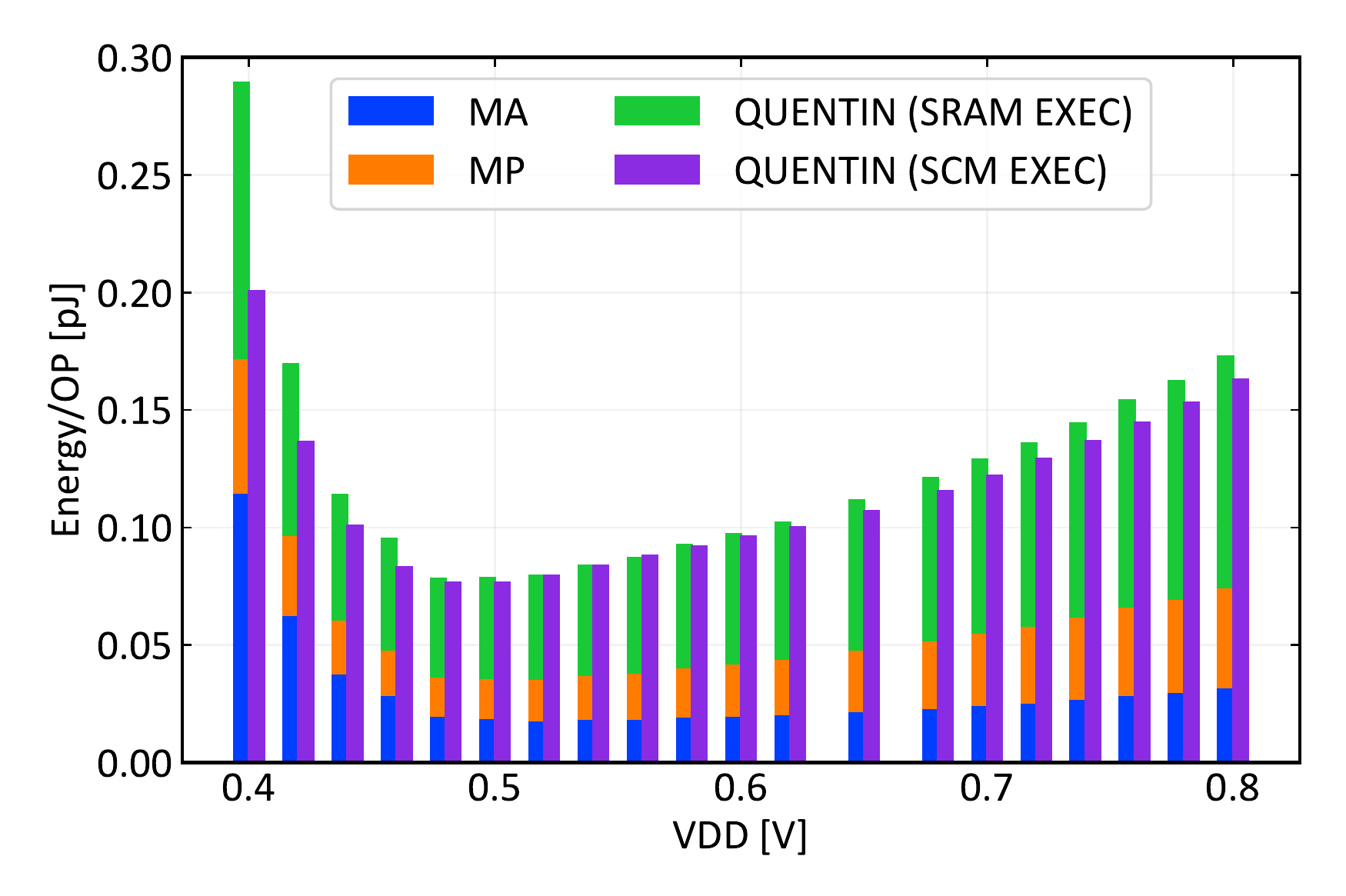}}
\caption{\textcolor{black}{SoC energy efficiency comparison when operating with supply voltage scaled SRAM, and when executing from SCM.}}
\label{fig:eff_comp}
\end{figure}

Overall, we observed that the biggest leakage contribution originates from the SRAM arrays, which are working 380~mV below the nominal specifications.
In HEFF mode (i.e. at a supply voltage of \SI{0.5}{\volt}), the SoC can sustain a clock frequency of 145~MHz, consuming \SI{2.5}{\milli \watt}. In this operating point, we report the highest energy efficiency achieved by the system, i.e. 12.7 Tops/s/W (49.2 Inference/s/mW). 
In HEFF operating mode, the leakage represents 32\% of the total power consumption.  \figurename \ref{fig:max_freq} reports the SoC maximum frequency used for the energy efficiency computation. Other relevant measurement points covering the entire operating range are reported in Table \ref{tab:raw-data}.

\figurename \ref{fig:eff_comp} shows the energy per binary operation when the SoC is executing either from SCMs only or SCMs plus SRAMs; The absolute lowest energy per binary operation is \SI{76}{\femto \joule \ OP}, which is achieved at \SI{0.46}{\volt} when executing from SCM only. Note that to execute a full neural network from SCM only is unrealistic since those memories are generally too small because of the low area density. The lowest energy per operation achievable when executing from SCMs and SRAMs is \SI{78}{\femto \joule \ OP}, and is reached at \SI{0.5}{\volt}. Overall, this plot demonstrates that our approach allows achieving comparable energy per operations both on SCMs and SRAMs. The aggressive voltage scaling performed in our experiments, together with a carefully crafted memory partition, significantly improves the energy efficiency when executing from dense SRAMs, ultimately relaxing the memory constraints for error-resilient applications deployment. The energy per operation reached at \SI{0.5}{ \volt} represents an improvement of 2.2X compared to the energy per operation measured at nominal condition \SI{170}{\femto \joule}.

Table \ref{tab:comparison} compares our work to BNN accelerator implementations that, similarly to our case, exploit BNN error-resilience to maximize energy efficiency. To our best knowledge, this work represents the first complete general purpose microcontroller architecture capable to exploit a heterogeneous memory hierarchy to execute error-resilient applications at the highest achievable energy efficiency.


\begin{figure}[t]
\centerline{\includegraphics[width=1\linewidth]{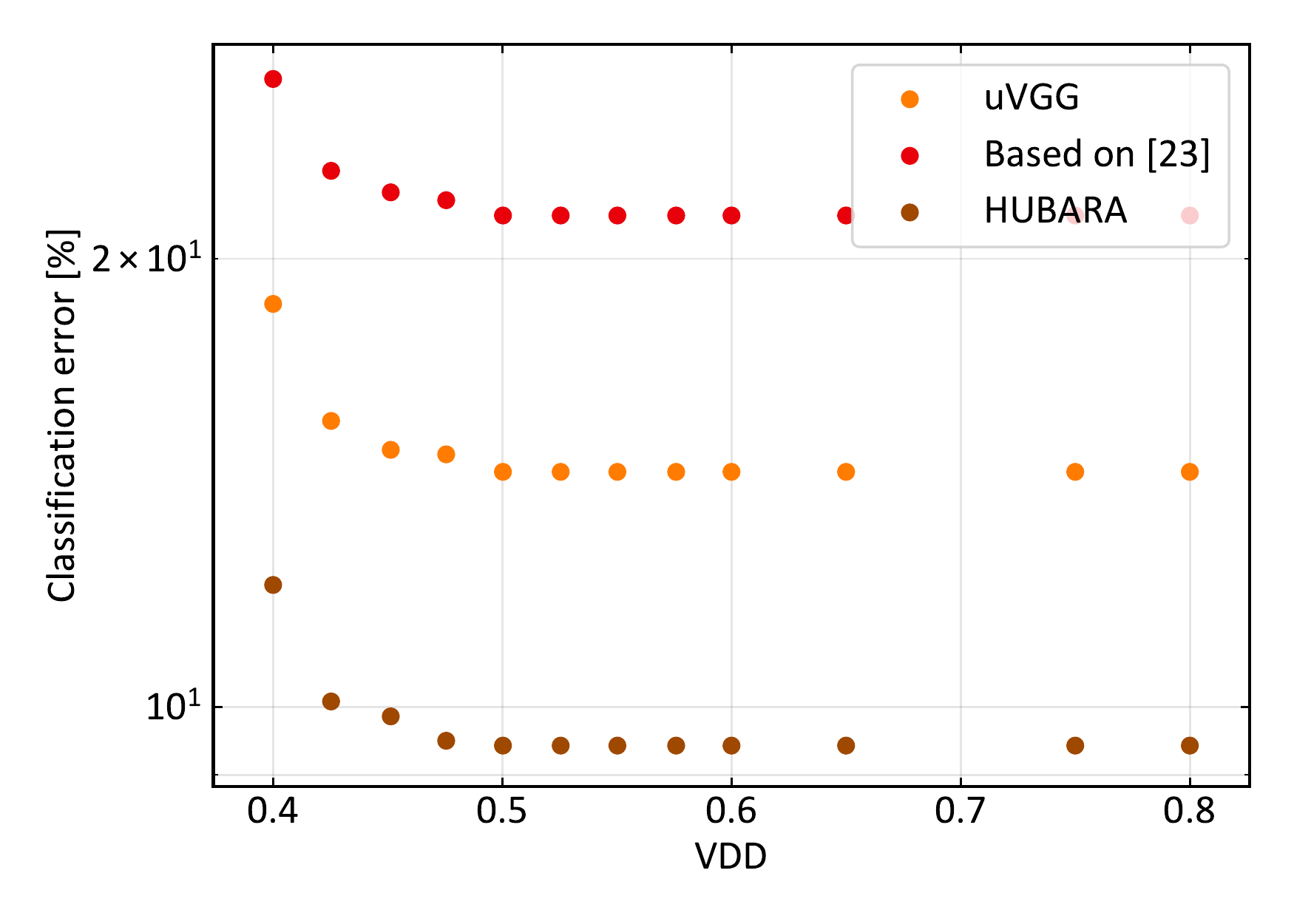}}
\caption{\textcolor{black}{SoC supply voltage / Accuracy tradeoff.}}
\label{fig:vdd_acc}
\vspace{-4mm}
\end{figure}

\begin{figure}[t]
\centerline{\includegraphics[width=1\linewidth]{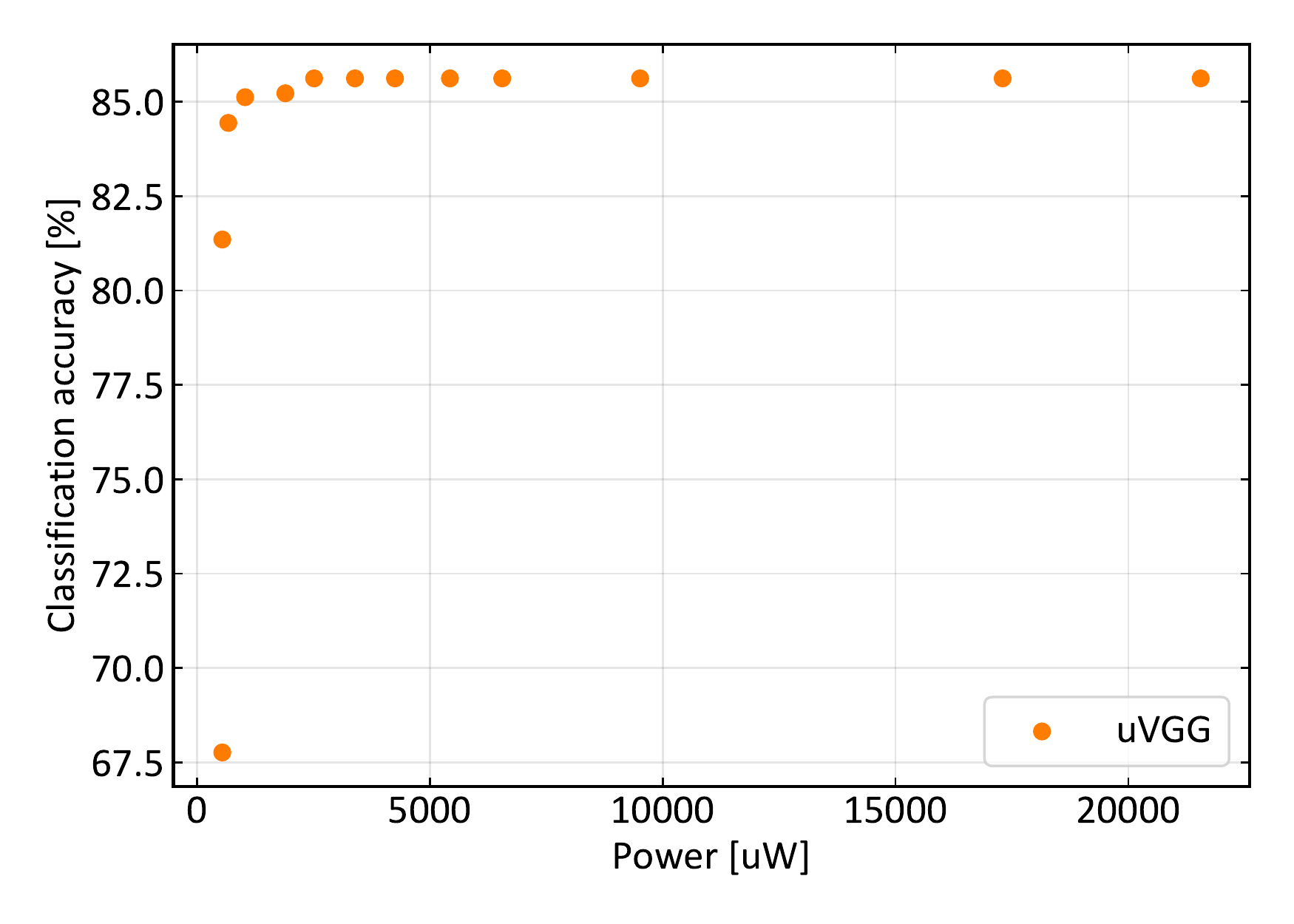}}
\caption{Power accuracy tradeoff evaluation for the uVGG BNN topology.}
\label{fig:power_acc}
\vspace{-4mm}
\end{figure}

\subsection{Power accuracy tradeoff}

From the analysis discussed in the previous section, we concluded that the final classification accuracy can be traded in spite of higher energy efficiency or lower power.
\figurename \ref{fig:vdd_acc} shows the accuracy loss versus the supply voltage on both memories and core logic. \figurename \ref{fig:power_acc} reports the accuracy loss versus the power consumption reduction enabled by the aforementioned supply voltage scaling. 

When the SoC operates in HEFF mode, we did not observe any accuracy loss. Thereby, we can conclude that, if performance (e.g. in terms of inference per second) is not a critical constraints, the energy efficiency can be improved by 2.2X (from \SI{170}{\femto \joule \ OP} to \SI{78}{\femto \joule \ OP}) without any appreciable penalty on the quality of the result, i.e. the classification accuracy of a BNN.

Furthermore, if a small classification accuracy loss can be tolerated by the application, smaller than 1\% in our analysis on typical target network topologies, the power consumption can be further pushed down, reducing it by 3.7X with respect to the HEFF operating mode. In this operating condition, the energy efficiency degrades compared to the HEFF operating point. This is caused by the fact that the total power consumption becomes leakage dominated as the supply voltage is reduced (\figurename \ref{fig:power_analysis}). Thereby, the performance reduction caused by the voltage scaling is not followed by a proportional power reduction. The ULP mode, where the chip consumes \SI{674}{\micro \watt}, is suitable for always-on operating scenarios, or IoT end-node with an expected lifetime in the order of months or years, as well as applications where the peak power dissipation is a critical concern (e.g. implantable devices).

\begingroup
\begin{table*}[t]
\footnotesize
\resizebox{\textwidth}{!}{%
\begin{tabular}{@{}lllllllllllll@{}}
\toprule
$V_{DD}$ & $F_{max}^{lin fit} $ & $P_{MA}^{tot}$ & $P_{MP}^{tot}$ & $P_{Quent.}^{tot}$ & $P_{MA}^{leak}$ & $P_{MP}^{leak}$ & $P_{Quent.}^{leak}$ & $P_{MA}^{dyn}$ & $P_{MP}^{dyn}$ & $P_{Quent.}^{dyn}$ & $Energy$ & $BER_{exp-fit}$ \\
{[}V{]} & {[}MHz{]} & \multicolumn{9}{c}{{[}uW{]}} & {[}pJ/OP{]} & {[}\%{]}\\
\midrule
0.42 & 18 & 247.5 & 135.2 & 292.1 & 226.4 & 105.4 & 211.7 & 21 & 29.8 & 80.4 & 0.290 & 0.001723 \\
0.46 & 56.9 & 353.1 & 244.9 & 598.1 & 272.9 & 127.6 & 265.2 & 80.2 & 117.3 & 332.9 & 0.114 & 0.000109 \\
0.5 & 145 & 589.2 & 544.5 & 1383 & 328.7 & 154.9 & 332.6 & 260.5 & 389.7 & 1050.3 & 0.079 & 6.93E-06 \\
0.54 & 205.1 & 823.5 & 839.5 & 2146.8 & 393.9 & 186.9 & 415.9 & 429.6 & 652.6 & 1730.9 & 0.080 & 4.40E-07 \\
0.58 & 265.3 & 1115.8 & 1219.8 & 3097.4 & 469.6 & 225.4 & 516.8 & 646.2 & 994.4 & 2580.6 & 0.087 & 2.79E-08 \\
0.62 & 325.4 & 1525.7 & 1792.2 & 4479.7 & 560.4 & 272.3 & 641.2 & 965.4 & 1519.8 & 3838.5 & 0.098 & 1.77E-09 \\
0.66 & 405.5 & 2022.3 & 2507.7 & 6167.2 & 670.6 & 330 & 797 & 1351.7 & 2177.7 & 5370.2 & 0.108 & 1.12E-10 \\
0.7 & 445.6 & 2473.1 & 3155.6 & 7616.3 & 799.6 & 399.6 & 993.5 & 1673.5 & 2756 & 6622.8 & 0.122 & 7.11E-12 \\
0.75 & 525.8 & 3075.6 & 4045.9 & 9622.4 & 953.5 & 480.8 & 1233.8 & 2122.1 & 3565.1 & 8388.6 & 0.136 & 4.51E-13 \\
0.80 & 565.8 & 3692 & 4945.1 & 11612.8 & 1137.9 & 582.6 & 1534.7 & 2554.1 & 4362.5 & 10078 & 0.154 & 2.86E-14 \\
\end{tabular}%
}
\caption{Raw data for frequency, power and energy at various voltage scaled operating points}
\label{tab:raw-data}
\end{table*}

\section{Conclusion}
\label{sec:conclusion}

In this paper, we presented a strategy to maximize energy efficiency in complex heterogeneous SoC. Our results demonstrated how to trade-off the energy consumption of an FDX 22nm SoC with the final classification accuracy of Binary Neural Networks, executed on a dedicated hardware accelerator. The proposed approach exploits the intrinsic noise robustness of BNN, i.e. the fact that a significant amount of noise on network parameters, quantified in terms of BER, marginally degrades the final classification accuracy. Our measurements show that thanks to a wise L2 memory partitioning, the system can operate reliably at very low voltages (i.e. down to \SI{0.42}{\volt}). Therefore, we show that by over scaling the supply voltage of the SRAMs of the SoC significantly below the nominal specifications, the energy per binary operation can be reduced by a factor of 2.2X compared to the nominal supply voltage. 
In this voltage over-scaled regime, we demonstrate that the reported energy efficiency gain does not affect the end-to-end classification accuracy of the BNN when the voltage is scaled down to \SI{0.5}{\volt}. Additionally, we show that, if a small penalty on the final classification accuracy is tolerable, e.g. within 1\%, the SoC can be operated in an ultra-low power mode, further reducing the overall power consumption (\SI{674}{\micro \watt} at \SI{18}{\mega \hertz}, \SI{0.42}{\volt}) without exceeding the energy consumption per binary operation shown at nominal operating conditions. 

\section{Acknowledgement}
This work was supported by the European H2020 FET project OPRECOMP (g.a. 732631)



\bibliographystyle{IEEEtran}
\bibliography{bibl.bib}

\begin{IEEEbiography}[{\includegraphics[width=1in,height=1.25in,clip,keepaspectratio]{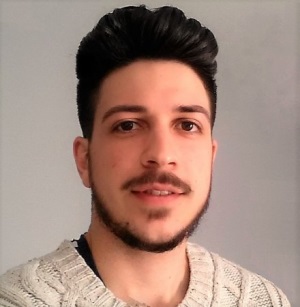}}]{Alfio Di Mauro}
received the M.Sc.  degrees in Electronic Engineering from the Electronics and Telecommunications Department (DET) of Politecnico di Torino in 2016. Since September 2017, he is currently pursuing the Ph.D. at the Integrated System Laboratory (IIS) of the Swiss Federal Institute of Technology of Zurich. His research focuses on the design of digital Ultra-Low Power (ULP) System-on-Chip (SoC) for Event-Driven edge computing.
\end{IEEEbiography}

\begin{IEEEbiography}[{\includegraphics[width=1in,height=1.25in,clip,keepaspectratio]{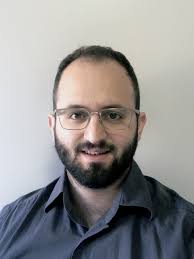}}]{Francesco Conti}
received the Ph.D. degree from the University of Bologna, Italy, in 2016. He is currently a Post-Doctoral Researcher  with the Integrated  Systems Laboratory, ETH Z\"urich, Switzerland, and also with the Energy-Efficient Embedded Systems Laboratory,  University  of  Bologna.  He  has  co-authored over  20  papers   on  international   conferences   and journals.  His  research  focuses  on  energy  efficient multicore architectures  and the applications  of deep learning  to low power digital  systems.
\end{IEEEbiography}

\begin{IEEEbiography}[{\includegraphics[width=1in,height=1.25in,clip,keepaspectratio]{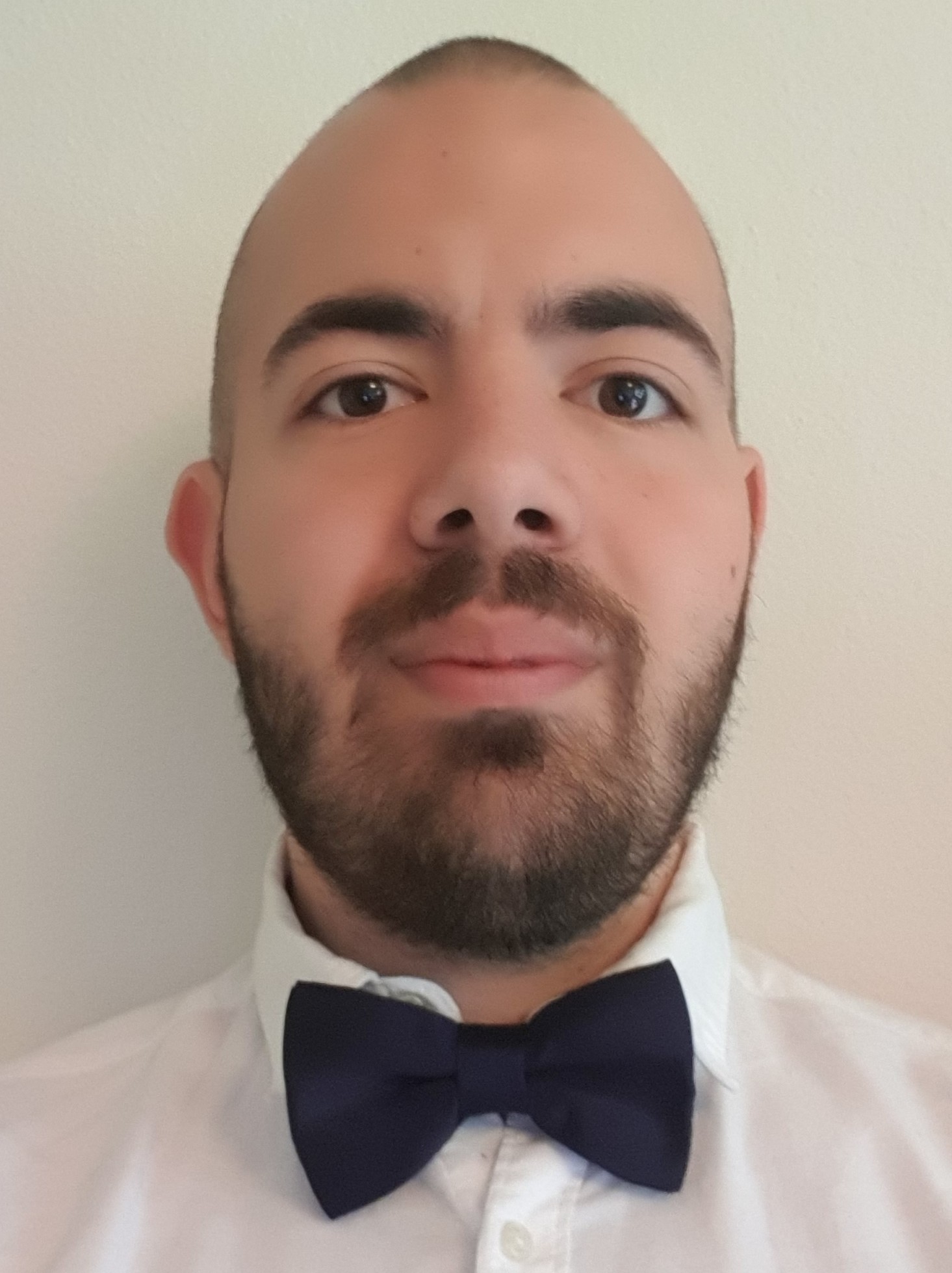}}]{Pasquale Davide Schiavone}
received the B.Sc. and M.Sc.  degrees  in  computer  engineering  from  the Polytechnic of Turin in 2013 and 2016, respectively.He   is   currently   pursuing   the   Ph.D.  degree   with the   Integrated   Systems   Laboratory,   ETH  Zürich.His   research   interests   include   low power   micro-processors  design  in  multi-core  systems  and  deep-learning  architectures  for energy-efficient  systems.
\end{IEEEbiography}

\begin{IEEEbiography}[{\includegraphics[width=1in,height=1.25in,clip,keepaspectratio]{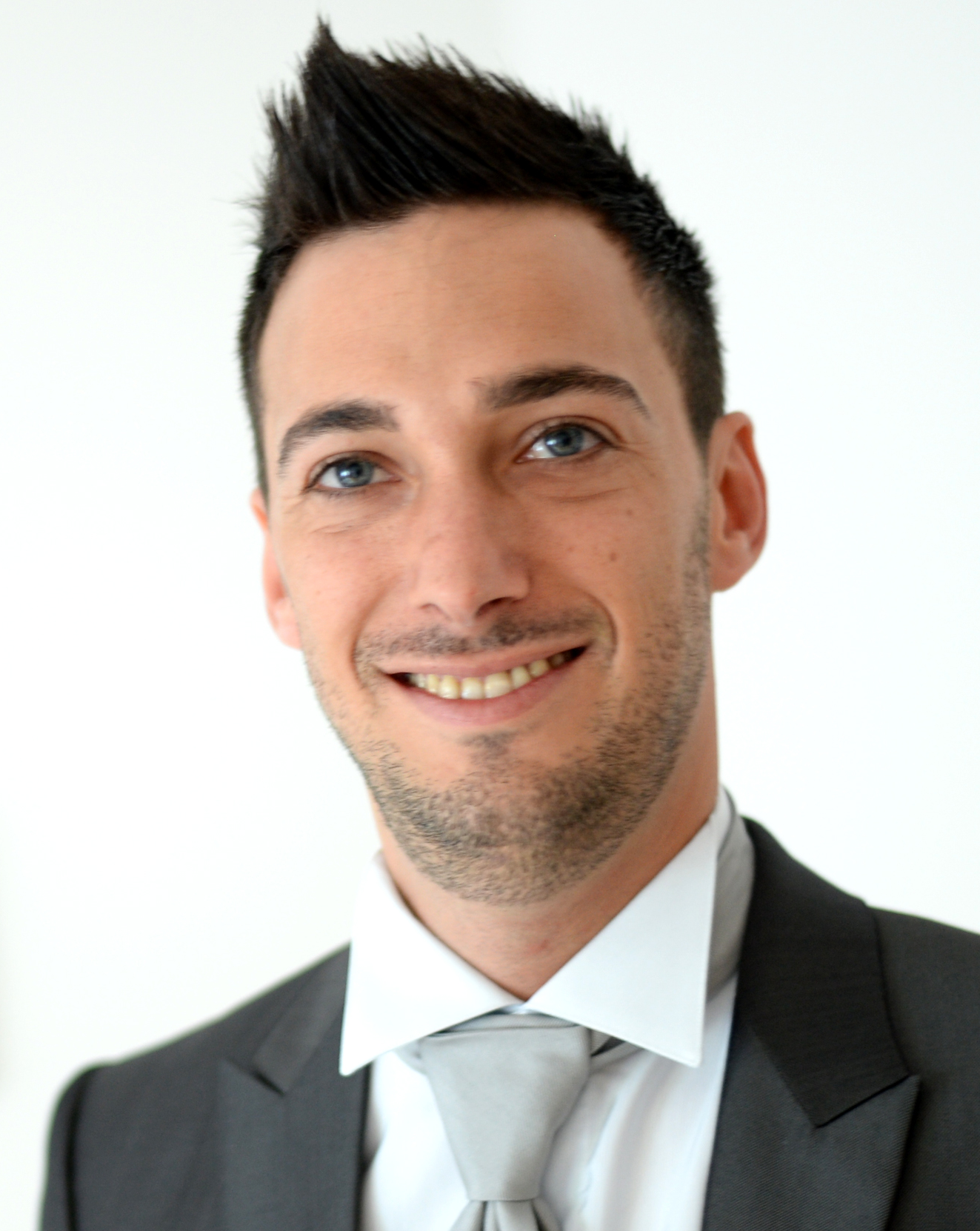}}]{Davide Rossi} received the PhD from the University of Bologna, Italy, in 2012. He has been a post doc researcher in the Department of Electrical, Electronic and Information Engineering
“Guglielmo Marconi” at the University of Bologna since 2015, where he currently holds an assistant professor position. His research interests focus on energy efficient digital architectures in the domain of heterogeneous and reconfigurable multi and many-core systems on a chip. This includes architectures, design implementation strategies, and runtime support to address performance, energy efficiency, and reliability issues of both high end embedded platforms and ultra-low-power computing platforms targeting the IoT domain. In
these fields he has published more than 100 papers in international
peer-reviewed conferences and journals. He is recipient of Donald O.
Pederson Best Paper Award 2018
\end{IEEEbiography}

\begin{IEEEbiography}[{\includegraphics[width=1in,height=1.25in,clip,keepaspectratio]{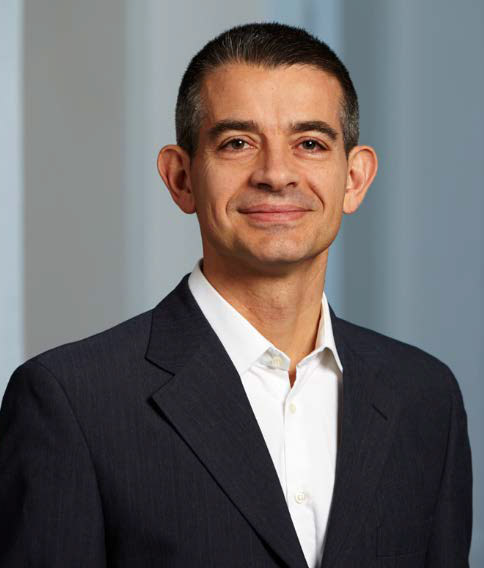}}]{Luca Benini}
received the Ph.D. degree in electrical engineering from Stanford University, Stanford, CA, USA, in 1997. He has served as the Chief Architect of the Platform 2012/STHORM Project with STMicroelectronics, Grenoble, France, from 2009 to 2013. He held visiting/consulting positions with École Polytechnique Fédérale de Lausanne, Stanford University, and IMEC. He is currently a Full Professor with the University of Bologna, Bologna, Italy. He has authored over 700 papers in peer-reviewed international journals and conferences, four books, and several book chapters. His current research interests include energy-efficient system design and multicore system-on-chip design. Dr. Benini is a member of Academia Europaea. He is currently the Chair of Digital Circuits and Systems with ETH Zürich, Zürich, Switzerland.
\end{IEEEbiography}

\end{document}